\documentclass{aa}  
\usepackage{color}
\usepackage{graphicx}
\usepackage{lscape}
\usepackage{natbib}
\usepackage{natbib,twoopt}
\usepackage[varg]{txfonts}
\usepackage{url}
\usepackage{xspace}
\usepackage{amssymb}
\usepackage{stmaryrd}
\usepackage{siunitx}
\usepackage{textcomp}

\bibpunct{(}{)}{;}{a}{}{,}

\def\etal{{et\,al.}\ }
\def\egb{\object{EGB\,6}}
\newcommand{\uc}{\object{UCAC2\,46706450}\xspace}
\newcommand{\uca}{\object{UCAC2\,46706450$-$A}\xspace}
\newcommand{\ucb}{\object{UCAC2\,46706450$-$B}\xspace}
\newcounter{Rco}

\newcommand{\logg}{\mbox{$\log g$}\xspace}
\newcommand{\loggw}[1]{\mbox{$\log g\hspace{-0.5mm} =\hspace{-0.5mm}  #1$}}

\newcommand{\Teff}{\mbox{$T_\mathrm{eff}$}\xspace}

\newcommand{\ebv}{$E_\mathrm{B-V}$\xspace}

\newcommand{\Lsol}{$L_\odot$}
\newcommand{\Msol}{$M_\odot$}
\newcommand{\Rsol}{$R_\odot$}

\begin{document}

\title{An extremely hot white dwarf with \\a rapidly rotating K-type subgiant companion: \uc
}

\author{Klaus Werner\inst{1} \and Nicole Reindl\inst{2} \and Lisa
  L\"obling\inst{1} \and Ingrid Pelisoli\inst{2} \and Veronika
  Schaffenroth\inst{2} \and \\Alberto Rebassa-Mansergas\inst{3,4} \and Puji
  Irawati\inst{5} \and Juanjuan Ren\inst{6}}

\institute{Institut f\"ur Astronomie und Astrophysik, Kepler Center for
  Astro and Particle Physics, Eberhard Karls Universit\"at, Sand~1, 72076
  T\"ubingen, Germany\\ \email{werner@astro.uni-tuebingen.de} 
\and
  Institut f\"ur Physik und Astronomie, Universit\"at Potsdam, Karl-Liebknecht-Stra\ss e 24/25, Germany
\and
Departament de F\'{i}sica, Universitat Polit\`{e}cnica de Catalunya, c/Esteve Terrades 5, E-08860 Castelldefels, Spain
\and
Institut d'Estudis Espacials de Catalunya, Ed. Nexus-201, c/Gran Capit\`{a} 2-4, E-08034 Barcelona, Spain
\and
National Astronomical Research Institute of Thailand, Sirindhorn AstroPark,
Donkaew, Mae Rim, Chiang Mai 50180, Thailand
\and 
Key Laboratory of Space Astronomy and Technology, National Astronomical
Observatories, Chinese Academy of Sciences, Beijing 100101, P. R. China
}

\date{3 June 2020 / 5 September 2020}

\authorrunning{K. Werner \etal}
\titlerunning{An extremely hot white dwarf with a rapidly rotating K-type subgiant companion}

\abstract{The subgiant \uc\ is a late-type star with an ultraviolet (UV) excess. It
  was considered as a candidate to establish a sample of stars of spectral type F, G, and K with
  white dwarf (WD) companions that could be used to test binary
  evolution models. To verify the WD nature of the companion, UV
  spectroscopy has previously been performed by other authors. Via a
  detailed model-atmosphere analysis, we show that the UV source is an
  extremely hot WD with an effective temperature of \Teff =
  $105\,000\pm5000$\,K, mass of $M/M_\odot = 0.54\pm0.02$, radius of
  $R/R_\odot = 0.040^{+0.005}_{-0.004}$, and luminosity of $L/L_\odot=
  176^{+55}_{-49}$, meaning that the compact object is just about to enter
  the WD cooling sequence. Investigating spectra of the cool star
  (\Teff = $4945\pm250$\,K), we found that it is a K-type subgiant with
  $M/M_\odot = 0.8-2.4$, $R/R_\odot = 5.9^{+0.7}_{-0.5}$, and
  $L/L_\odot= 19^{+5}_{-5}$ that is rapidly rotating with $v
  \sin(i)=81$\,km\,s$^{-1}$. Optical light curves reveal a period of
  two days and an o-band peak-to-peak amplitude of 0.06\,mag. We
  suggest that it is caused by stellar rotation in connection with
  star spots. With the radius, we infer an extremely high rotational
  velocity of $v_{\mathrm{rot}}=151^{+18}_{-13}$\,km\,s$^{-1}$, thus
  marking the star as one of the most rapidly rotating subgiants
  known. This explains chromospheric activity observed by H\,$\alpha$
  emission and emission-line cores in \ion{Ca}{ii} H and K as well as
  NUV flux excess. From equal and constant radial velocities of the WD
  and the K subgiant as well as from a fit to the spectral energy
  distribution, we infer that they form a physical, wide (though
  unresolved) binary system. Both components exhibit similar metal
  abundances and show iron-group elements with slightly oversolar (up
  to 0.6~dex) abundance, meaning that atomic diffusion in the WD
  atmosphere is not yet active due to a residual, weak
  radiation-driven wind. Kinematically and from its height above the
  Galactic plane, the system belongs to the Galactic thick disk,
  indicating that it is an old system and that the initial masses of both
  stars were close to 1\,$M_\odot$.}


\maketitle
%

\section{Introduction}
\label{sect:intro}

Stellar multiplicity is an omnipresent outcome of the star-formation
process.  About every other solar-type (mass $M\approx0.7-1.3$\,\Msol)
star is found in the binary system (see \citealt{Duchene+Kraus2013} for a
review). At solar metallicity, about one quarter of these stars is
found in close binaries (orbital period $P<10^4$\,d, separation
$a<10$\,AU), and the close-binary fraction is found to increase
strongly with decreasing metallicity and may be as high as 53\% at a
metallicity of [Fe/H] $= -3$~dex \citep{2019ApJ...875...61M}.  When
the more massive member of the binary evolves off the main sequence,
such close systems will eventually interact with each other by
exchanging mass and angular momentum \citep{2004A&A...419.1057W}. This
in turn influences their evolution and can lead to a broad range of
astrophysical phenomena that are absent for the life of single
stars. Systems that will undergo a common-envelope event will end up
as very close binaries with final orbital periods typically between
0.1 and 10\,d \citep{2011A&A...536A..43N}, or even merge
\citep{2002MNRAS.336..449H}.  Systems that transferred mass via stable
Roche-lobe overflow or wind accretion can be found at longer final
periods of a few $10^{2}$--$10^{3}$\,d \citep{1989A&A...214..186P,
  2002MNRAS.336..449H, 2012MNRAS.423.2764N, 2013MNRAS.434..186C}.
Evolved systems with periods longer than about 3000\,d are thought to
always remain detached \citep{2017A&A...597A..68V}.

Detailed studies of evolved binaries are fundamental for various
reasons. The mass and period distribution of very short orbital period
binaries provides important observational constraints on the poorly
understood common-envelope phase \citep{1986A&A...168..105R,
  2010A&A...513L...7S, 2010MNRAS.403..179D, 2011A&A...536L...3Z}. In
addition, these binaries can be employed to search for and study
supernovae type Ia progenitor candidates \citep{Napiwotzkietal2001,
  2013A&A...554A..54G, SG+2015, Rebassa-Mansergas+2019, Reindl+2020}
or help us to understand how binary interactions alter
the intrinsic properties of the stars (such as their atmospheric
composition, rotational rates, pulsations, mass loss, dust formation,
and circumstellar-envelope morphology;
\citealt{2018arXiv180900871V}). Last but not least, binaries that
avoided mass exchange can be used to investigate the initial-final
mass relation, which is a key constraint on stellar evolution theory
and important to understanding the chemical enrichment and the efficiency
of star formation in galaxies \citep{2005AJ....129.2428S,
  2008A&A...477..213C, 2012ApJ...746..144Z, 2014MNRAS.440.3184B,
  2015ApJ...815...63A}.

Aiming to provide a large sample to test binary evolution models and
type Ia supernovae formation channels, \citet{2016MNRAS.463.2125P}
established a group of 934 main-sequence FGK stars from the Large Sky
Area Multi-Object Fiber Spectroscopic Telescope (LAMOST,
\citealt{Zhao+2012}) survey and the Radial Velocity Experiment survey
\citep{2013AJ....146..134K}, which show excess flux at ultraviolet
(UV) wavelengths, and hence likely have a white dwarf (WD) companion.
Such systems are still very rarely known in comparison to thousands of
M stars with WD companions \citep{2013MNRAS.435.2077H, 2016MNRAS.458.3808R}.
For nine objects in their sample, \citet{2016MNRAS.463.2125P} obtained follow-up
spectroscopy with the \emph{Hubble Space Telescope} (HST).

One of these nine systems, \uc, is the subject of this work. A
spectroscopic analysis of the late-type star was performed by various
authors revealing, for example, an effective temperature of \Teff
$=4905\pm16$\,K and a surface gravity of \logg $=2.90\pm0.04$
\citep{Ho+2017}, indicating that the star is an early K subgiant.
Assuming that the observed Ly\,$\alpha$ line in the HST spectrum is of
photospheric origin, a model-atmosphere fit by
\citet{2016MNRAS.463.2125P} to the Ly\,$\alpha$ profile yielded \Teff
= 24\,000\,K and a low surface gravity of \logg $\approx 5.0$,
indicating that the hot component is a pre-WD object like a hot
subdwarf.

In the present paper, we analyze in detail the UV spectrum of the
compact companion and show that its temperature was strongly
underestimated, and that it is among the hottest known WDs (Sect.\,\ref{sect:observations}). We then reassess the spectroscopic
observations to characterize the K-type star
(Sect.\,\ref{sect:coolstar}), search for radial velocity (RV)
variations (Sect.\,\ref{sect:rvv}), and investigate its Galactic
population membership (Sect.\,\ref{sect:distance}). We derive stellar
parameters of both components (Sect.\,\ref{sect:radii}) and examine
the optical photometric variability (Sect.\,\ref{sect:lightcurve}). We
conclude in Sect.\,\ref{sect:conclusions}. 

In the following, we refer to the spectroscopic binary components as \uca\ and \ucb\ for the K subgiant and the hot WD, respectively.

\begin{figure*}[t]
 \centering  \includegraphics[width=1.0\textwidth]{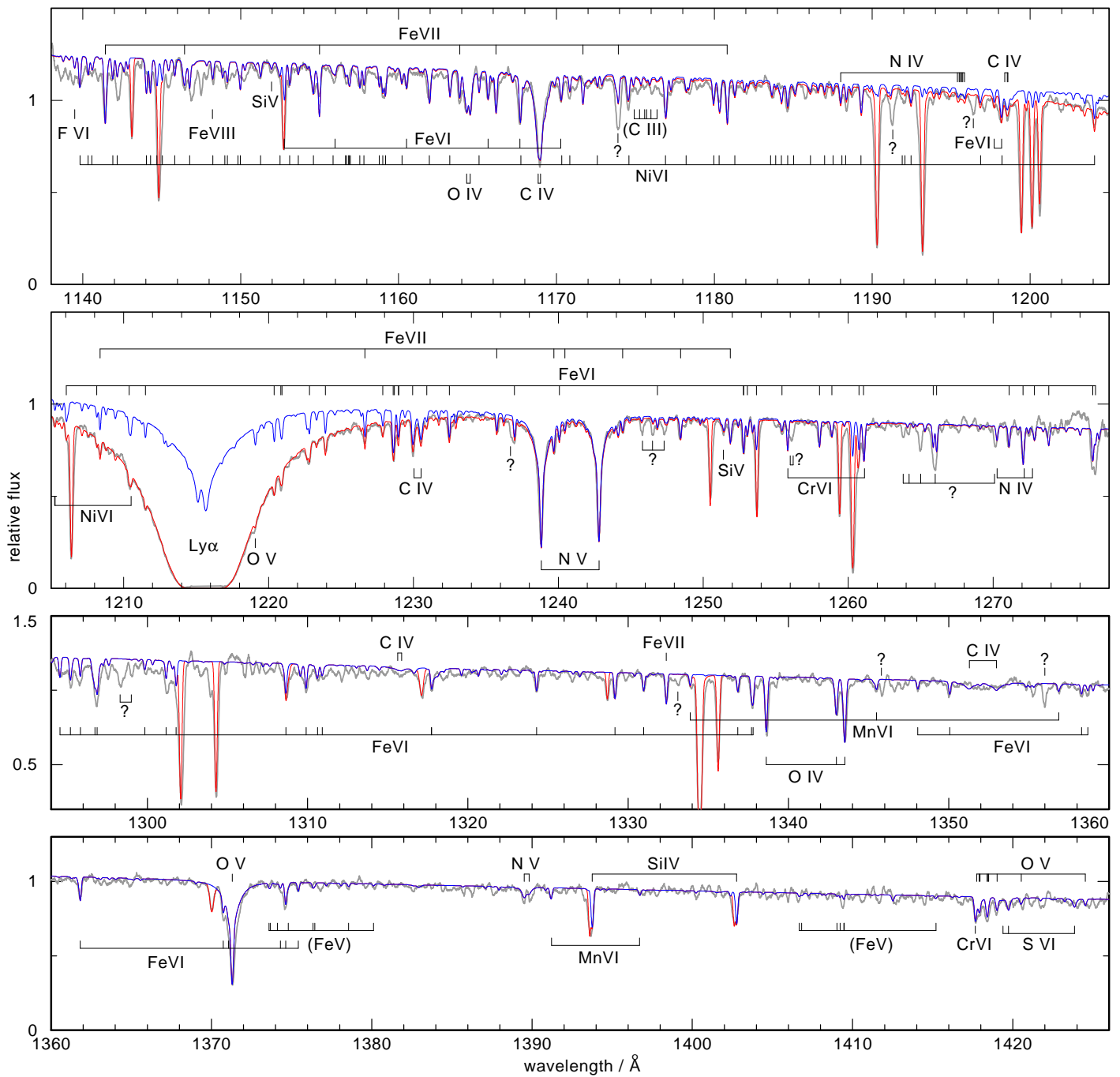}
  \caption{HST/COS spectrum of the white dwarf in \uc\ (gray) compared
    to a photospheric model spectrum (blue graph: \Teff = 105\,000\,K,
    \logg = 7.4) with the measured metal abundances
    (Tab.\,\ref{tab:results}) and assuming solar H and He
    abundances. The same model attenuated by interstellar lines is
    plotted in red. Prominent photospheric lines are
    identified. Identifications in brackets denote uncertain
    detections. Question marks indicate unidentified photospheric
    lines.}
\label{fig:ucac_hst}
\end{figure*}

\begin{figure*}[t]
 \centering  \includegraphics[width=1.0\textwidth]{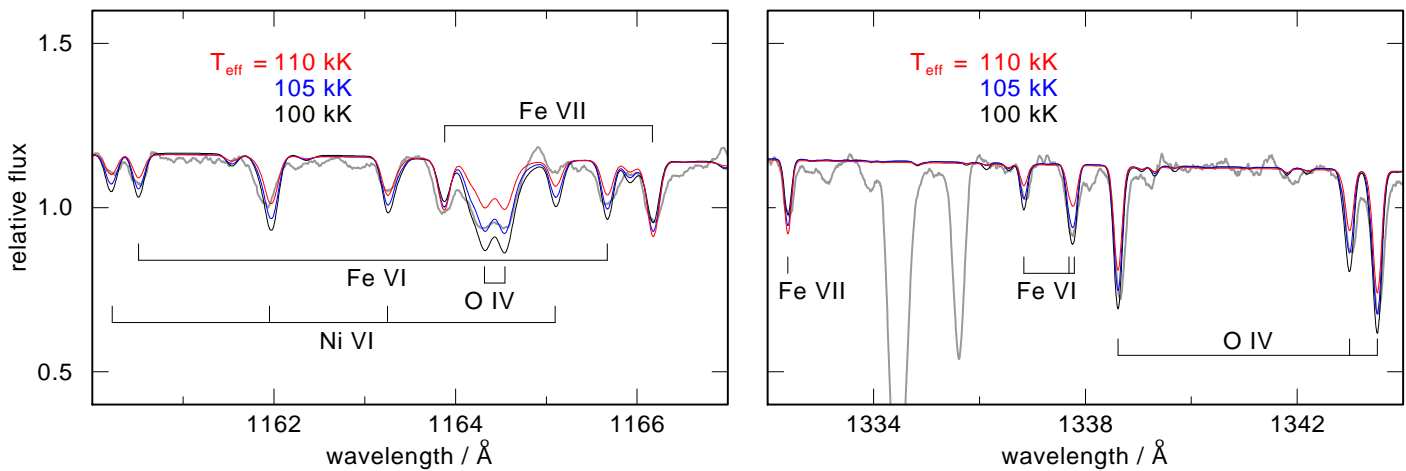}
  \caption{Details of observed WD spectrum compared with three
    models with different temperatures (\Teff =
    $105\,000\pm5000$\,K) representing the error range in determination of
    \Teff. \emph{\emph{Left:}} the \ion{O}{iv} multiplet gets weaker with
    increasing \Teff, as well as lines from \ion{Fe}{vi} and
    \ion{Ni}{vi}, while \ion{Fe}{vii} lines get stronger. \emph{\emph{Right:}} same
    behavior of oxygen and iron lines in another wavelength range.
}
\label{fig:teff}
\end{figure*}

\begin{figure}[t]
 \centering  \includegraphics[width=0.6\columnwidth]{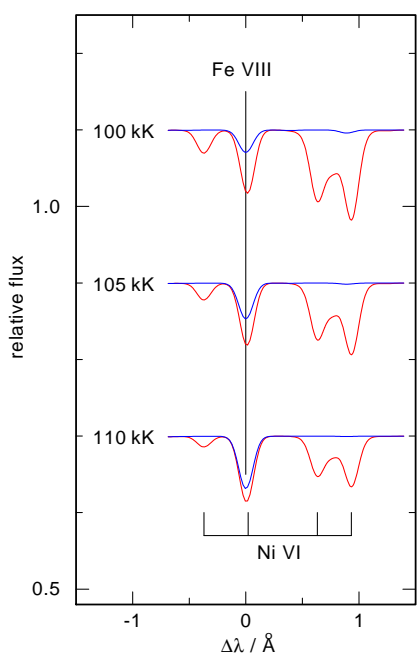}
  \caption{Effect of increasing temperature on the \ion{Ni}{vi}/\ion{Fe}{viii} line blend at 1148.2\,\AA\ (\Teff=100,
    105, and 110\,kK, from top to bottom). Blue graphs show spectra
    computed without Ni lines, that is to say, we only see the \ion{Fe}{viii}
    line profile. The nickel lines become weaker and the iron line
    becomes stronger. As a result, the line Fe+Ni blend stays almost
    constant. The models have abundances of Fe = $3.5 \times 10^{-3}$ 
and Ni =$3.0 \times 10^{-4}$ by mass.}\label{fig:feviii}
\end{figure}

\begin{figure*}[t]
 \centering  \includegraphics[width=0.95\columnwidth]{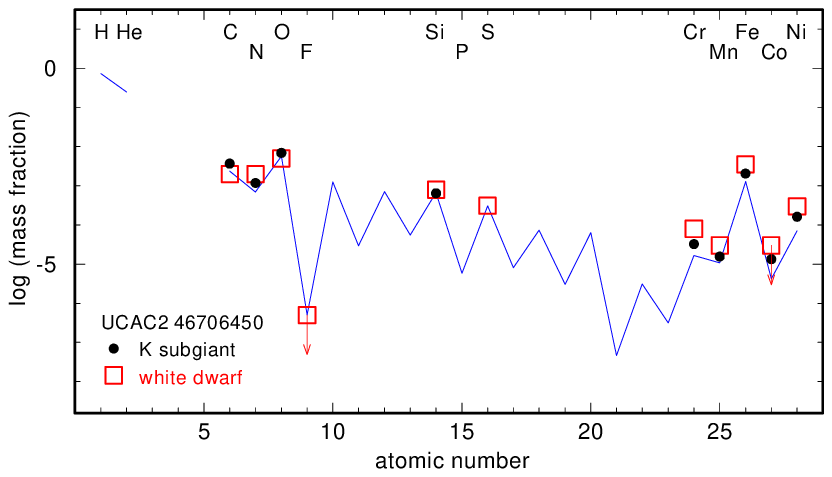}\hspace{5mm}
 \centering  \includegraphics[width=0.95\columnwidth]{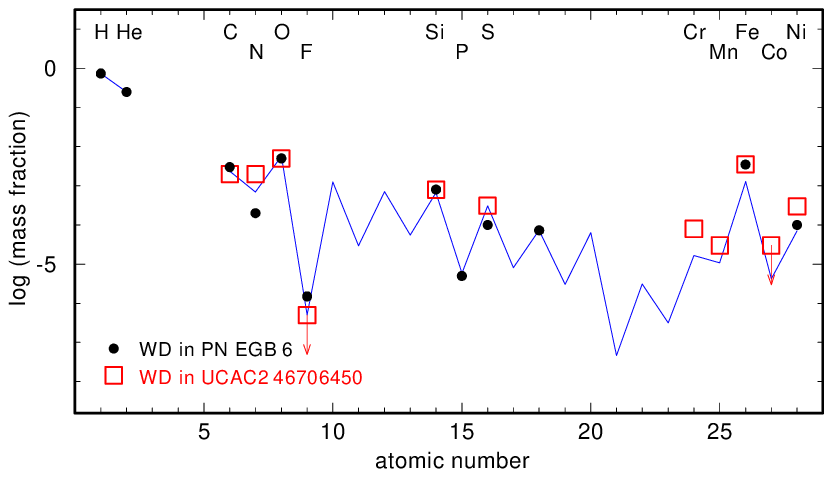}
  \caption{\emph{\emph{Left:}} element abundances measured in the WD (red
    squares) and in the K subgiant (black dots) of the binary
    \uc. \emph{\emph{Right:}} abundances in the WD of \uc\ (red squares),
    compared to the DAO WD in \egb\ (black dots). The blue line
    indicates solar abundances.}\label{fig:abu}
\end{figure*}

\section{Analysis of UV spectroscopy of the hot white dwarf}
\label{sect:observations}

The star \uc\ was observed on Oct. 30, 2014, with the Cosmic Origins
Spectrograph (COS) aboard HST, using the
G130M grating centered on 1291\,\AA\ for a 2109\,s exposure (dataset
LCKY08010, PI: S. Parsons). The spectrum was retrieved from the MAST
archive. The approximate useful wavelength coverage is 1138--1278\,\AA\ and
1296--1426\,\AA. The spectral resolution is about 0.1\,\AA. For our
analysis, we smoothed the spectra with a 0.1\,\AA-wide boxcar to
increase the signal-to-noise ratio (SNR). For comparison with observations,
model spectra were convolved with a Gaussian (FWHM = 0.14\,\AA). The
spectrum and our model fit are shown in
Fig.\,\ref{fig:ucac_hst}. We identified lines from highly ionized
light metals, namely, \ion{C}{iv}, \ion{N}{iv-v}, \ion{O}{iv-v},
\ion{Si}{iv-v}, \ion{S}{vi} as well as from iron-group elements,
namely, \ion{Cr}{vi}, \ion{Mn}{vi}, \ion{Fe}{vi-viii}, and
\ion{Ni}{vi}.  Evidently the spectrum is that of a very hot WD.
Concerning the ionization stages observed in the UV, it appears
similar to other hot objects, for example, the DA PG\,0948+534 with \Teff =
105\,000\,K \citep{2019MNRAS.483.5291W} and the DAO central stars of
Sh\,2-216 and \egb\ with \Teff = 95\,000\,K and 105\,000\,K,
respectively \citep{2007A&A...470..317R,2018A&A...616A..73W}. 

The broad Ly\,$\alpha$ profile is dominated by interstellar
hydrogen, and the COS spectrum does not cover any helium line. Thus,
the H/He ratio is unknown, and in principle the object could be either a
He-rich (DO or DOA) WD or a H-rich (DA or DAO) WD.
Nevertheless, we show that the effective temperature and
metal abundance measurements are rather independent of this.

The photospheric lines are blueshifted by $-10$\,km\,s$^{-1}$. Accordingly, the
observed spectrum is shifted to rest wavelengths in all figures
presented here. Some of the most prominent interstellar lines
were modeled. They are blueshifted between $-15$ and $-35$\,km\,s$^{-1}$. From
Ly\,$\alpha,$ we derived a neutral H column density of $\log n_H =
20.4\pm 0.04$ toward the WD. Comparing the continuum shape of our
final model and observation, we found a reddening of \ebv$= 0.03\pm
0.01$. The models presented here are attenuated by this reddening
value. We note the reddening derived by us also agrees with
  values reported in the 2D dust maps of \cite{Schlegel1998} and
  \cite{Schlafly2011}, as well as with the lower limit provided
  by the 3D dust map of \cite{Lallement+2018}.

We used the T\"ubingen Model-Atmosphere Package
(TMAP\footnote{\url{http://astro.uni-tuebingen.de/~TMAP}}) to compute
non-LTE, plane-parallel, line-blanketed atmosphere models in radiative
and hydrostatic equilibrium
\citep{1999JCoAM.109...65W,2003ASPC..288...31W,tmap2012}. The models
include H, He, C, N, O, Si, P, S, Cr, Mn, Fe, Co, and Ni. The employed
model atoms are described in detail by \citet{2018A&A...609A.107W}. In
addition, we performed line formation iterations (i.e., keeping the
atmospheric structure fixed) for fluorine using the model atom
presented in \citet{2015A&A...582A..94W}.

\subsection{Effective temperature, surface gravity, and H/He ratio}
\label{sect:teff}

We started our analysis by fixing the value of the surface gravity to
\logg = 7.4. At \Teff $\approx$ 100\,000\,K, this corresponds to a WD
with a mass of $\approx$ 0.54\,M$_\sun$. We also fixed the H/He ratio to the
solar value. We then varied \Teff\ and the metal abundances to obtain
a good fit to the line features. We found \Teff = $105\,000 \pm
5000$\,K and the abundances listed in Table~\ref{tab:results}. The
abundance measurements are described in detail below
(Sect.\,\ref{sect:metals}). Finally, we looked how \logg\ can be
constrained and how the H/He ratio affects the derived atmospheric
parameters.

The relatively small error in \Teff\ follows from the fact that we can
exploit several ionization balances, first of all, iron. We do not see
\ion{Fe}{v}, imposing a strict lower limit of 100\,000\,K. The
relative strengths of \ion{Fe}{vi} to \ion{Fe}{vii} lines
then yield the value of 105\,000\,K (Fig.\,\ref{fig:teff}). In
addition, at temperatures above 110\,000\,K, the only observed
\ion{Fe}{viii} line (at 1148.2\,\AA) becomes too strong (a problem with
a blending \ion{Ni}{vi} line is discussed in
Sect.\,\ref{sect:metals}). Another sensitive indicator for \Teff\ is
oxygen. With increasing temperature, the \ion{O}{iv} multiplets
(Fig.\,\ref{fig:teff}) and \ion{O}{v}~1371\,\AA\ become
weaker, whereas the highly excited \ion{O}{v} multiplet at
1420\,\AA\ becomes stronger. The detection of \ion{N}{iv} lines (around
1190\,\AA\ and 1271\,\AA) rules out a \Teff\ value well over
105\,000\,K. The absence of the \ion{C}{iii} multiplet at 1175\,\AA,\ on
the other hand, requires \Teff\ above 100\,000\,K.

Next, we looked at how the surface gravity affects the model
spectrum. Reducing \logg\ from the fiducial value 7.4 to 7.1 results
in unobserved sharp line cores of the \ion{C}{iv} doublet at
1352\,\AA. An increase to \logg = 7.7 results in too broad wings of
the \ion{C}{iv} lines observed at 1178\,\AA\ and 1198\,\AA. We  adopt
\logg = $7.4 \pm 0.5$, with a conservative error estimate.

We then explored how the metal lines depend on different values of
the H/He ratio, which remains undetermined at the moment. The
reduction of helium from the assumed solar value down to $1.3 \times
10^{-5}$ (mass fraction) is affecting the UV metal line strengths only
marginally. The increase of helium to 98\,\% is slightly shifting the
ionization balance of metals to lower stages such that an increase of
\Teff\ by about 5000\,K would be necessary to match the observed line
strengths. Therefore, within error ranges, the measured metal
abundances do not depend on the assumed H/He ratio and hold
irrespective of whether the star is a DA or a DO white dwarf.

\subsection{Metal abundances}
\label{sect:metals}

We present in detail how
elemental abundances were inferred. The employed lines are identified in
Fig.\,\ref{fig:ucac_hst}. The resulting abundance values are listed in
Tab.\,\ref{tab:results} and displayed in Fig.\,\ref{fig:abu}.

\subsubsection{CNO}

Three \ion{C}{iv} multiplets are detectable and were fit, namely
3d--4f, 3d--4p, and 3p--4s, at 1169\,\AA, 1198\,\AA, and 1230\,\AA,
respectively. It is worthwhile to note that two other, more highly
excited, \ion{C}{iv} multiplets are not detectable: 4p--7d at
1316\,\AA\ and the blend of 4d--7f and 4f--7g at 1351\,\AA\ and
1353\,\AA. They serve to constrain the upper limit on the C
abundance and the lower limit on surface gravity (see
Sect.\,\ref{sect:teff}).

Besides the strong \ion{N}{v} resonance doublet, we observe a weak and
highly excited \ion{N}{v} multiplet (4s--5p) at 1390\,\AA, and several
\ion{N}{iv} lines. There is a singlet at 1188\,\AA\ ($^1$S--$^1$P$^{\rm
  o}$), which, however, is blended by a \ion{Ni}{vi} line and a
multiplet at 1196\,\AA\ ($^3$D--$^3$P$^{\rm o}$). A triplet appears at
1271\,\AA\ ($^3$P$^{\rm o}$--$^3$D), but two components are blended by
\ion{Fe}{vi} lines.

Two prominent \ion{O}{iv} multiplets are detected, namely a blend of
components at 1164\,\AA\ ($^2$D--$^2$F$^{\rm o}$) and the three
components of the $^2$P--$^2$D$^{\rm o}$ transition at 1338.6\,\AA,
1343.0\,\AA, and 1343.5\,\AA. We also detect several \ion{O}{v} lines:
the prominent \ion{O}{v} 1371\,\AA, the $^1$D--$^1$D$^{\rm o}$ singlet
at 1219\,\AA,\ and an accumulation of weak lines in the region
1418--1426\,\AA, which stem from the transitions $^3$P$^{\rm
  o}$--$^3$D and $^3$D--$^3$F$^{\rm o}$. 

\begin{table}[t]
\begin{center}
\caption{Atmospheric parameters of the WD in
  \uc.\tablefootmark{a} }
\label{tab:results} 
\begin{tabular}{rrr}
\hline 
\noalign{\smallskip}
\Teff/\,K & $105\,000 \pm 5000$ \\
$\log$($g$ / cm\,s$^{-2}$) & $7.4 \pm 0.5$      \\
\noalign{\smallskip}
\hline 
\noalign{\smallskip}
abundance   & $X_i$               & \qquad[$X_i$]\\
\hline 
\noalign{\smallskip}
C         & $2.0 \times 10^{-3}$ & $-$0.07 \\ 
N         & $2.0 \times 10^{-3}$ & 0.46 \\ 
O         & $5.0 \times 10^{-3}$ & $-$0.06 \\ 
F         & $<5.0 \times 10^{-7}$ & $<0.16$ \\ 
Si        & $8.0 \times 10^{-4}$ & 0.08 \\ 
S         & $3.1 \times 10^{-4}$ & 0.00 \\ 
Cr        & $8.0 \times 10^{-5}$ & 0.70 \\
Mn        & $3.0 \times 10^{-5}$ & 0.45 \\
Fe        & $3.5 \times 10^{-3}$ & 0.46 \\ 
Co        & $<3.0 \times 10^{-5}$ & $<0.91$ \\
Ni        & $3.0 \times 10^{-4}$ & 0.64 \\ 
\noalign{\smallskip} \hline
\end{tabular} 
\tablefoot{  \tablefoottext{a}{Abundances given in mass
    fractions (column 2) and logarithmic abundances relative to solar
    value \citep[column 3; solar abundances
      from][]{2009ARA&A..47..481A}. Error limits for the abundances
    are $\pm 0.5$~dex.}  } 
\end{center}
\end{table}

\subsubsection{Light metals: F, Si, and S}

We detect a very weak feature at the position of the
\ion{F}{vi}~1139.5\,\AA\ line, which is often observed in hot WDs
\citep[e.g., \egb,][]{2018A&A...616A..73W}. It can be fit with a
solar fluorine abundance, however, this result is regarded as
uncertain given the poor SNR at the blue edge of the
COS spectrum. We therefore adopted the solar abundance as an upper limit.

Besides the \ion{Si}{iv} resonance doublet at 1394/1402\,\AA, we
detect two weak \ion{Si}{v} lines at 1152.0\,\AA\ and 1251.4\,\AA,
which are components of the $^3$P$^{\rm o}$--$^3$P and $^3$P$^{\rm
  o}$--$^3$D transitions, respectively. The \ion{Si}{iv} doublet is
blended by an ISM component that is blueshifted by $-25$\,km\,s$^{-1}$
relative to the photospheric one. A weak \ion{S}{vi} line at
1419.7\,\AA\ is identified, which is the strongest of three components
of the 4d--5p transition, allowing for a sulfur abundance
measurement.

\subsubsection{Iron group elements: Cr, Mn, Fe, Co, and Ni}

A line of \ion{Cr}{vi} is visible at 1417.7\,\AA. Another one, at
1261.1\,\AA,\ is a blend with an \ion{Fe}{vi} line. A third one is seen
at 1255.8\,\AA, but this is blended with a broad, unidentified
feature.
Five weak \ion{Mn}{vi} lines are detected at 1333.9\,\AA,
1345.5\,\AA, 1356.9\,\AA, 1391.2\,\AA, and 1396.7\,\AA.

As mentioned previously, we identified iron lines from
\ion{Fe}{vi-viii}. Care must be taken when using the only
\ion{Fe}{viii} line (at 1148.2\,\AA) as a temperature indicator
(Fig.\,\ref{fig:feviii}). Its strength increases with \Teff, but there
is a blending \ion{Ni}{vi} line whose strength decreases with
\Teff. In effect, the line blend stays constant in the temperature
regime discussed here. We do not see lines of \ion{Fe}{v}. Some of the
\ion{Fe}{v} lines that we see in models at slightly lower temperature
in the wavelength region of 1373--1414\,\AA\ are indicated in
Fig.\,\ref{fig:ucac_hst}. 

Our models predict many \ion{Co}{vi} lines, mainly in the
short-wavelength range of the COS spectrum, provided the cobalt
abundance is high enough. We do not detect any such line and derive an
upper abundance limit.

Many \ion{Ni}{vi} lines are found, mainly in the short wavelength
region up to about 1210\,\AA. Similarly to iron, \ion{Ni}{v} lines are
absent due to the high effective temperature.

\subsubsection{Unidentified lines}

A number of photospheric lines remain unidentified. The
strongest appear at these wavelengths: 1173.6, 1191.0, 1195.7,
1236.4, 1246.2, 1256.3, 1266.9, 1298.3, 1332.8, 1345.5, and
1355.7\,\AA. These are indicated in Fig.\,\ref{fig:ucac_hst} with a
question mark. We also found these lines in other hot WDs or central
stars of planetary nebulae (PNe) and suspect that some of them could be
unknown \ion{Fe}{vii} lines.

\subsubsection{Summary of abundance analysis of the WD}

Within error limits, the abundances of light metals metals are solar
(Tab.\,\ref{tab:results}, Fig.\,\ref{fig:abu}). For the iron-group
elements a slight enhancement (0.45--0.70~dex oversolar) is
found. This overabundance might point to radiative acceleration acting
in the atmosphere, but we report below that the K-subgiant
also exhibits an overabundance of these elements (left panel of
Fig.\,\ref{fig:abu}). Hence, the element abundance pattern probably
reflects the initial metallicity of both stars. In the right panel of
Fig.\,\ref{fig:abu}, we compare the abundances found for \ucb\ with
those of the WD central star of the PN \egb, which has the same
\Teff\ and \logg\ \citep{2018A&A...616A..73W}. Apart from nitrogen,
the abundance patterns are rather similar.

\section{Optical and IR spectroscopy of the K star}
\label{sect:coolstar}

Subgiant \uca\ dominates the flux in the optical and infrared (IR) wavelength
range. Optical spectroscopy ($\lambda \approx 3700-9000$\,\AA) of the
system was obtained in course of the LAMOST survey (resolving power
$R\approx 1800$). Furthermore, moderately high-resolution ($R\approx
22\,500$) IR ($\lambda \approx 15\,100-17\,000$\,\AA) spectroscopy was
obtained with the Apache Point Observatory Galactic Evolution
Experiment (APOGEE; \citealt{Apogee2017}). In the course of the Gaia mission,
a medium-resolution ($R \approx 11\,700$) spectrum over the wavelength
range 8450--8720\,\AA\ (around the \ion{Ca}{ii} triplet) was collected
with the Radial Velocity Spectrometer.

Additional follow-up spectroscopy were obtained at the 5.1-meter
Palomar Hale telescope in California, USA, and the Thai National
Telescope (TNT) at the National Observatory in Thailand. One spectrum
was obtained at the Palomar Hale Telescope during the night of the
Jan. 20, 2014. We used the double spectrograph together with
the 1200 lines/mm grating in the red and the 600 lines/mm grating in
the blue. The observations were carried out using the long slit with 1\arcsec\ width.
This configuration resulted in a wavelength coverage of
$\approx7600-9300$\,\AA\ in the red and $\approx3500-6500$\,\AA\ in
the blue at $R=6400$ and $R=1400$, respectively. The spectrum was
reduced and calibrated using the pamela \citep{1989PASP..101.1032M}
and molly\footnote{Tom Marsh's molly package is available at
  \url{http://deneb.  astro.warwick.ac.uk/phsaap/software}} softwares,
respectively. Thirteen additional spectra were obtained during the 2015--2017 period at
the TNT using the Middle Resolution
fiber-fed Echelle Spectrograph (MRES). The spectra covered the
$\approx4300-8800$\,\AA\ wavelength range at $R=15\,000$. The spectra
were reduced and calibrated using the DECH software package.

The two LAMOST spectra from data release 2 (DR2) were analyzed by
\cite{Luo+2016}, who derived \Teff, \logg, [Fe/H], as well as the
RV values from the individual observations. These
spectra were also analyzed by \cite{Ho+2017}, who provide results for
\Teff, \logg, [Fe/H], as well as the $\alpha$-enhancement
[$\alpha$/M]. Two additional LAMOST spectra were obtained within the
DR5 that were analyzed (along with the previous spectra) by
\cite{Luo+2019}. Abundances of several elements (C, Cl, N, O, Mg, Al,
Si, P, K, Ca, Ti, V, Cr, Mn, Fe, Co, Ni, and Rb), \Teff, \logg, and RV
from the APOGEE spectrum are provided by the APOGEE Stellar Parameter
and Chemical Abundances Pipeline (ASPCAP, \citealt{ASPCAP+2016}). Some
of these element abundances relevant for the comparison with the WD
companion are depicted in the left panel of Fig.\,\ref{fig:abu}. The
iron-group elements are slightly enhanced, by 0.2--0.5~dex.
An interesting point to notice is that ASPCAP predicts an unusually high
rotational velocity of $v \sin(i)=80.76$\,km\,s$^{-1}$. Atmospheric parameters 
from the APOGEE spectrum were also derived with the data driven method
called the Cannon \citep{Ness+2015, Casey+2016}. However, the $\chi^2$
of the Cannon fit is quite large, indicating that the parameters and
abundances are not reliable, and thus we did not consider these values
further. Finally, the Gaia DR2 also provides values for \Teff\ and
RV.

\begin{table}[t]
\caption{Atmospheric parameters, masses as obtained from evolutionary
  calculations, and radii and luminosities of the subgiant \uca and the white dwarf \ucb.}
\begin{tabular}{l l l} 
\hline 
\noalign{\smallskip}
       & \small\uca  & \small\ucb \\
\noalign{\smallskip}
\hline 
\noalign{\smallskip}
\Teff / K       & $4945 \pm 250$    & $105\,000 \pm 5000$ \\
\noalign{\smallskip}
$\log$($g$ / cm\,s$^{-2}$)  & $3.04\pm0.25$ & $7.4\pm0.5$ \\
\noalign{\smallskip}
$M$ / \Msol     & $0.8-2.4$         & $0.54\pm0.02$ \\
\noalign{\smallskip}
$R$ / \Rsol     & $5.9^{+0.7}_{-0.5}$ & $0.040^{+0.005}_{-0.004}$  \\
\noalign{\smallskip}
$L$ / \Lsol     & $19^{+5}_{-5}$     & $176^{+55}_{-49}$ \\
\noalign{\smallskip}
\hline 
\end{tabular}
\label{tab:RLM}
\end{table}

\begin{figure*}[ht]
\centering
\includegraphics[width=\textwidth]{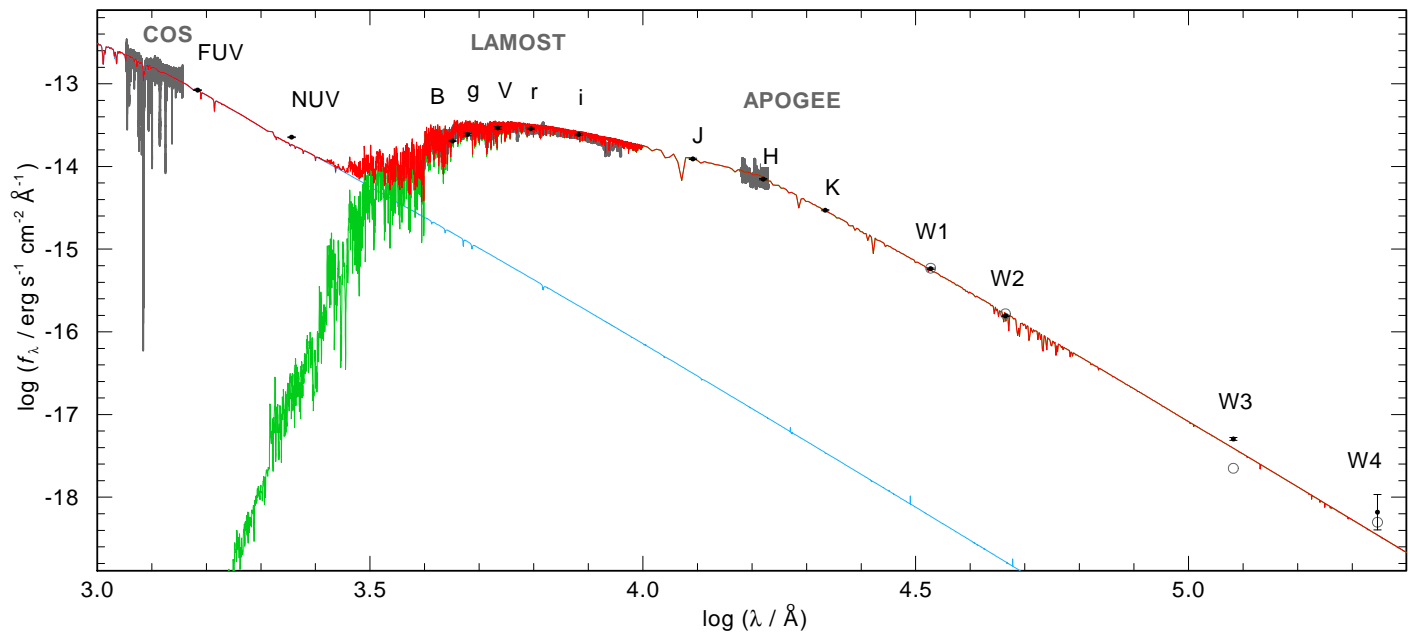}
\caption{Fit to SED of \uc. HST/COS, LAMOST (normalized to r-band
  magnitude), and APOGEE spectra are shown in gray. The model fluxes
  from the hot WD and the K-type subgiant are shown in light
  blue and green, respectively. The combined model flux is shown in
  red. The black dots indicate filter-averaged fluxes that were
  converted from observed magnitudes. Gray open circles indicate WISE
  fluxes for which the flux correction was not taken into account.}
\label{fig:sed}
\end{figure*}

In Table\,\ref{tab:RLM}, we provide the average of the \Teff\ and \logg\
values derived by the various analyses and give an estimate of the error,
which also includes the systematic error (estimated from the scatter of
the results from the individual analyses). We note that the fits to the
optical spectrum do not consider the contribution of the flux of the
WD to the blue part of the spectrum (see Fig.\,\ref{fig:sed}). On the
other hand, values from the fit to the LAMOST spectrum do not differ
too much from what is obtained from the APOGEE and Gaia spectra.

\begin{figure}[t]
\centering
\includegraphics[width=0.8\columnwidth]{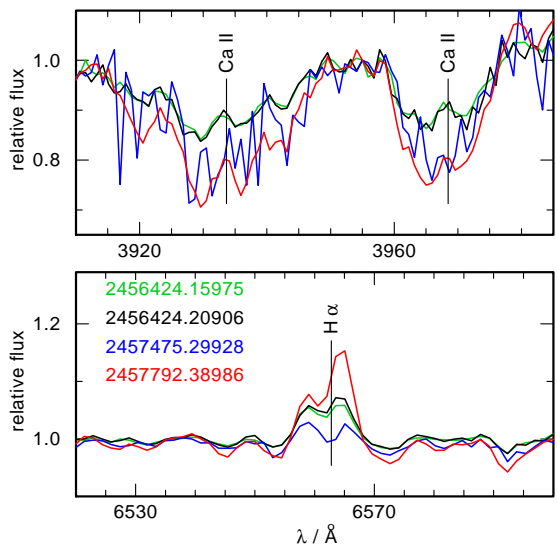}
\caption{Time-variable emission line cores in the \ion{Ca}{ii} H and K
  doublet and H\,$\alpha$ emission observed in the LAMOST spectra
  indicate that \uca is chromospherically active. The HJD of the
  observations is provided.}
\label{fig:emission}
\end{figure}

\begin{figure}[t]
\centering
\includegraphics[width=0.8\columnwidth]{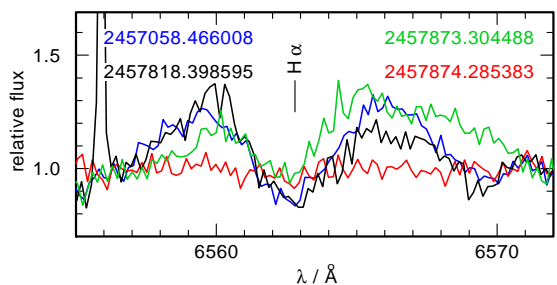}
\caption{Time-variable H\,$\alpha$ emission observed in TNT spectra
  indicate that \uca is chromospherically active. The HJD of the observations
  is provided.}
\label{fig:emission2}
\end{figure}

A closer inspection of the LAMOST spectra reveals that the line cores
of the \ion{Ca}{ii} H and K doublet (Fig.\,\ref{fig:emission}) appear
in emission. The double-peaked H\,$\alpha$ emission (also seen in the
TNT spectra, Fig.\,\ref{fig:emission2}), can be explained by the
superposition of a very broad emission line plus a photospheric
absorption line. The strength of these emission lines is found to be
time variable. Emissions in these lines are typical for
chromospherically active stars \citep{Wilson1963,
  Wilson1968,GrayCorbally2009}.

\section{Radial velocity variability}
\label{sect:rvv}

\begin{table}[t]
\begin{center}
\caption{HJD and radial velocities of \uca as measured from the different spectra.}
\label{tab:rvs} 
\begin{tabular}{cll}
\hline 
\noalign{\smallskip}
HJD & $v_{\mathrm{rad}}$\,[km\,s$^{-1}$]        &       survey  \\
\noalign{\smallskip}
\hline 
\noalign{\smallskip}
2456424.148720  & $     -22.0   \pm     10.6    $ &     LAMOST  \\
2456424.160200  & $     -23.7   \pm     10.6    $ &     LAMOST  \\
2456424.171710  & $     -21.6   \pm     10.6    $ &     LAMOST  \\
2456424.190910  & $     -22.9   \pm     10.5    $ &     LAMOST  \\
2456424.210070  & $     -20.3   \pm     10.5    $ &     LAMOST  \\
2456424.228520  & $     -18.6   \pm     10.6    $ &     LAMOST  \\
2456678.061937  & $     -21.9   \pm     6.0     $ &     Palomar \\
2456709.957320  & $     -18.81318       \pm     0.15    $ &     APOGEE  \\
2456717.996280  & $     -17.94056       \pm     0.12    $ &     APOGEE  \\
2456735.937290  & $     -19.22651       \pm     0.17    $ &     APOGEE  \\
2457058.466008  & $     -27.2   \pm     3.4     $ &     TNT     \\
2457475.290870  & $     -26.0   \pm     10.4    $ &     LAMOST  \\
2457475.300180  & $     -26.0   \pm     10.4    $ &     LAMOST  \\
2457475.309490  & $     -21.5   \pm     10.5    $ &     LAMOST  \\
2457792.389120  & $     -31.5   \pm     13.2    $ &     LAMOST  \\
2457817.394727  & $     -33.1   \pm     3.4     $ &     TNT     \\
2457818.398595  & $     -21.0   \pm     3.4     $ &     TNT     \\
2457819.380124  & $     -18.1   \pm     4.1     $ &     TNT     \\
2457819.400669  & $     -30.3   \pm     7.3     $ &     TNT     \\
\noalign{\smallskip} \hline
\end{tabular} 
\end{center}
\end{table}

The scatter of the RV values from the three individual APOGEE spectra (which
were all obtained within one month) is smaller than 1\,km\,s$^{-1}$,
which already indicates that the system is very likely not a close
binary. We also searched for RV variations using the values we measured
from the LAMOST, Palomar, and TNT spectra, as well as those provided
in the catalogs for the APOGEE and Gaia spectra (total time coverage
is about four years, see Table\,\ref{tab:rvs}). In order to check if the RV variations are
significant or are merely produced by random fluctuations, we followed
the approach as outlined in \cite{Maxted2001} and
\cite{Geier+2015}. For this, we calculated the inverse-variance weighted
mean velocity from all RV measurements. Assuming this mean velocity to
be constant, we calculated the $\chi^2$. Comparing this value with the
$\chi^2$ distribution for the appropriate number of degrees of
freedom, the probability $p$ of obtaining the observed value of
$\chi^2$ or higher from random fluctuations around a constant value
can be calculated. We obtain a value of $\log p=-1.8,$ and hence
conclude that \uca is not significantly RV variable and very likely
not a close binary.\footnote{\cite{Geier+2015} considered objects with
  $\log p>4.0$ as significantly variable, and objects with $-1.3 >
  \log p > -4.0$ as possibly variable.} 

It is worth mentioning that \cite{2017MNRAS.472.4193R}
claimed \uca to be RV variable. They obtained the RV values by fitting the
\ion{Na}{i} absorption doublet at around 5900\,\AA. However, revising these
fits, we realized that some of the RVs they used were affected by
embedded emission (likely resulting from the night sky) and are hence
not reliable. By excluding these RV values, \uca does not show RV
variation, in agreement with the results obtained here.

The WD spectrum is blueshifted by $-10\pm
8$\,km\,s$^{-1}$. Considering the gravitational redshift of
$v_\mathrm{grav} = 8.6\pm1.3$\,km\,s$^{-1}$ gives a RV value for the WD of
$v_\mathrm{rad} = -18.6\pm8.1$\,km\,s$^{-1}$ (calculated using the mass
and radius from Sect.\,\ref{sect:radii}). This agrees very well
with the most precisely measured RV value of the K star
($v_\mathrm{rad} = -18.35\pm0.07$\,km\,s$^{-1}$ from ASPCAP) and can
be seen as a first hint that the system is indeed a physically connected
binary and that the velocities of both stars represent the system
velocity.

We stress, however, that with the current set of RV measurements we
are only able to detect RV amplitudes larger than $\approx
15$\,km\,s$^{-1}$, and that in case of a low inclination angle or a
long orbital period, the amplitude can be much smaller than this. 

\section{Distance and Galactic population membership}
\label{sect:distance}

The distance of \uca (the dominant source in the optical) based on the
Gaia DR2 parallax is provided in the \cite{Bailer-Jones+2018} catalog
and amounts to $d=1417^{+45}_{-43}$\,pc. From this we calculate a height
above the Galactic plane of $z=1154^{+36}_{-35}$\,pc, that is to say we find
that it lies in a region where the thick disk dominates.

The Galactic population membership of \uc can be further constrained
by looking at the chemistry of \uca, as well as the kinematics of the
system. Both thick disk and thin disk stars are observed over a wide
metallicity range ([Fe/H] between $-0.8$ and 0.2~dex,
\citealt{Kordopatis+2015}), but thick disk stars are found to be more
$\alpha$-rich at a given metallicity compared to thin disk stars (e.g.,
\citealt{Fuhrmann1998, Recio-Blanco+2014}). At metallicities above
[Fe/H] $=0.1$~dex, the separation of the two populations is, however,
unclear (e.g., \citealt{Hayden+2017}).
At the average values [Fe/H] $= 0.09$~dex (standard deviation $0.10$~dex)
and [$\alpha$/M] $=0.02$~dex (standard deviation $0.07$~dex) reported for
\uca in the LAMOST and ASPCAP catalogs, the star appears just on
the borderline between the two populations (see, e.g., Fig.\,1 in
\citealt{Hayden+2017}). Therefore, it is worth also looking at the
kinematics of the system. Since LAMOST, APOGEE, TNT, Palomar, Gaia, and HST/COS
spectra do not indicate significant RV variations, we adopt the mean
RV ($-21.95\pm4.55$, where the error represents the standard deviation of
the individual measurements) as system velocity. With this and the proper
motion and parallax from the Gaia DR2, we calculate space velocities of
$U=46.4\pm1.0$\,km\,s$^{-1}$, $V=-21.8\pm2.5$\,km\,s$^{-1}$, and
$W=-23.6\pm3.7$\,km\,s$^{-1}$(\footnote{Cardinal directions, with U
    being positive toward the Galactic center. Values have been adjusted for
    the Solar motion using the numbers of \cite{Tian+2015}.}).
The kinematic Galactic population membership can then be derived on the
basis of the Mahalanobis distance, which is the number of standard deviations
between the space velocities of \uc and the velocity ellipsoids of the
thin disk, thick disk, and the halo \citep{Gianninas+2015}. Using the
velocity ellipsoid values from \cite{Kordopatis+2011}, we find that the
kinematics of \uc clearly point to membership of the thick disk.  A
final, and probably most decisive, criterion that can be used to
determine the Galactic population membership is the age of the system,
and this is discussed further in the next section.

\section{Radii, luminosities, masses, and ages}
\label{sect:radii}

In order to determine the radii of the two stellar components, we
performed a fit to the spectral energy distribution (SED), by varying
their solid angles $\pi(R/d)^2$ (which relates the astrophysical flux
on the surface of the stars to what is received on Earth) until a good
agreement of the combined synthetic flux and the observations was
found. We assumed a reddening of \ebv$=0.03$ (see
Sect.\,\ref{sect:observations}) and the distance provided by
\cite{Bailer-Jones+2018} based on the Gaia parallax. For \ucb, we used
our best fit model from Sect.\,\ref{sect:observations}. For \uca, we used
a NextGen model \citep{Allard+2012} with \Teff$=4950$\,K, \loggw{3.0}, and
a metallicity of [Fe/H] = 0~dex, which is close to the average values of
the derived parameters from various analyses.
The errors on the radii were determined taking into account the
uncertainties of the effective temperatures and the
distance given by \cite{Bailer-Jones+2018}.

\begin{figure}[t]
\centering
\includegraphics[width=\columnwidth]{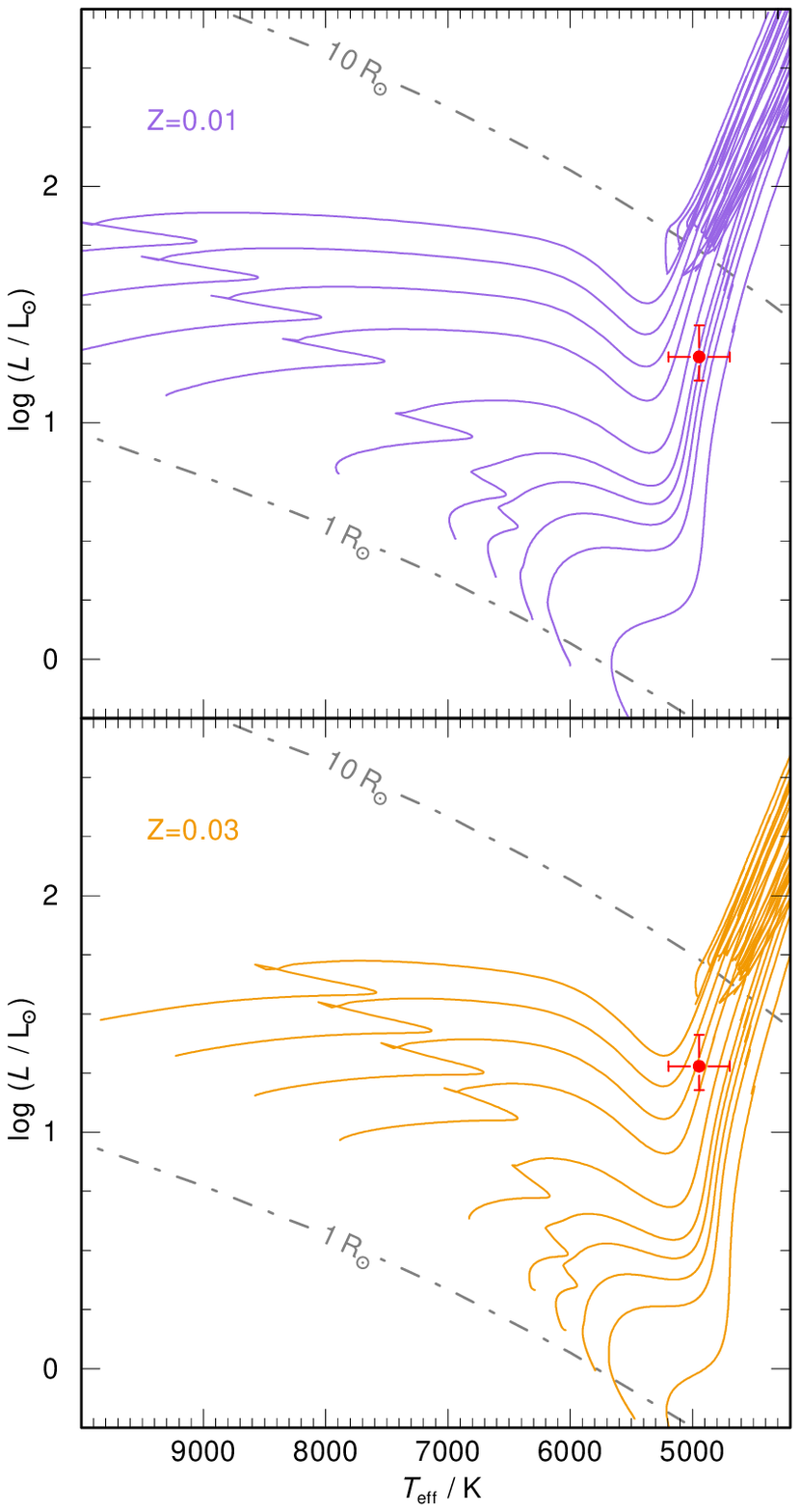}
\caption{Location of \uca\ (red) in the HRD compared to stellar
  evolutionary tracks of \cite{Pietrinferni+2004} for initial masses of 0.8, 1.0,
  1.1, 1.2, 1.3, 1.5, 1.8, 2.0, 2.2, and 2.4\,\Msol. 
  The upper panel indicates evolutionary tracks calculated for $Z=0.01$
  ([Fe/H] = 0.25~dex) and a He mass fraction of 0.259. The lower panel shows
  tracks calculated for $Z=0.03$ ([Fe/H] = 0.26~dex) and a He
  mass fraction of 0.288. The gray dashed-dotted
  lines indicate radii of 1 and 10\,$R_\odot$.}
\label{fig:HRD}
\end{figure}

Our best fit is shown in Fig.\,\ref{fig:sed}. The black dots indicate
filter-averaged fluxes that were converted from observed
magnitudes. GALEX $FUV$ and $NUV$ magnitudes were taken from
\cite{Bianchi2014} and converted to fluxes as outlined in
\cite{Reindl+2016}. $B$, $V$, $g$, $r$, and $i$ magnitudes were taken
from \cite{Henden+2015}, 2MASS $J$, $H$, and $K$ magnitudes from
\cite{Cutri+2003}. To convert these magnitudes into fluxes, we used
the zero-points provided in \cite{HolbergBergeron2006}.
\emph{Wide-field Infrared Survey Explorer} (WISE) $W1$, $W2$, $W3$, and $W4$
magnitudes were taken from \cite{Cutri+2003} and converted to fluxes
using the zero-points and the flux corrections for a K2V star provided
in \cite{Wright+2010}. The light blue, green, and red lines in
Fig.\,\ref{fig:sed} indicate the SEDs of \ucb, \uca, and the combined
best fit model, respectively. All model fluxes are corrected for
interstellar reddening, using the reddening law of
\cite{fitzpatrick1999} with \ebv$=0.03$. The gray lines in
Fig.\,\ref{fig:sed} correspond to the HST/COS, co-added
LAMOST, and APOGEE spectra. Overall, the combined model flux
reproduces the observed SED well. Only the observed NUV flux exceeds 
the predicted flux by about 0.1~dex. We attribute this to the chromospheric flux of
the K-type star, as chromospheric fluxes are known to exceed the photospheric flux
by several orders of magnitude in chromospherically active stars
\citep{2013MNRAS.431.2063S,2016MNRAS.463.1844S,Dixon+2020}. Based on the
NUV excess of 0.1~dex, we estimate that chromospheric flux of the K-type star
exceeds its photospheric flux by about two orders of magnitude and which is also
what is found for other such rapidly rotating and chromospherically active
giants \citep{Dixon+2020}. The WISE $W3$ and $W4$ fluxes also indicate a
slight excess, but these bands are in particular sensitive to the flux
corrections (gray open circles in Fig.\,\ref{fig:sed} indicate WISE fluxes for
which the flux correction was not taken into account). Thus, we regard the
far-IR excess as uncertain.

\begin{figure}[t]
\centering
\includegraphics[width=\columnwidth]{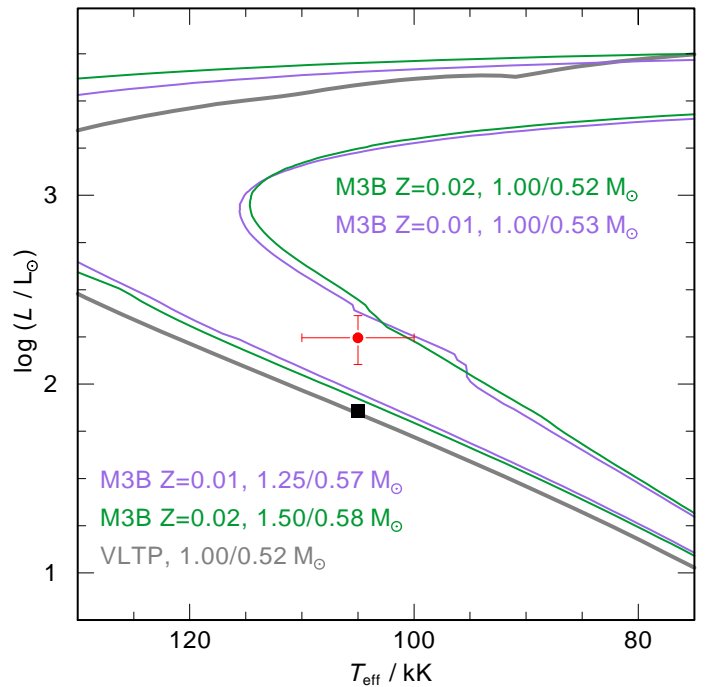}
\caption{Location of \ucb\ (red) in the HRD compared to stellar
  evolutionary tracks of \cite{MillerBertolami2016} calculated for
  $Z=0.01$ (purple) and $Z=0.02$ (green). Initial and final masses of
  the tracks are indicated. Shown in gray is a VLTP evolutionary
  track. The black square indicates the position of \object{EGB\,6}.}
\label{fig:HRD2}
\end{figure}

For \uca, we derive a radius of $R_A=5.9^{+0.7}_{-0.5}$\,\Rsol,\ and for
the hot WD, $R_B=0.040^{+0.005}_{-0.004}$\,\Rsol. Using $L=4\pi\,\sigma
R^2 T_\mathrm{eff}^4$, where $\sigma$ is the Stefan-Boltzmann
constant, we calculate luminosities of $L_A=19^{+5}_{-5}$\,\Lsol\ and
$L_B=176^{+55}_{-49}$\,\Lsol\, for \uca and \ucb, respectively. These
values are what is expected for a subgiant star and a hot WD that is
just about to enter the WD cooling sequence. Hence, we conclude that
the system is indeed a physical binary.

With the spectroscopically determined surface gravities (average value
for the K star from Table\,\ref{tab:RLM}) and using $M=g R^2/G$, where
$G$ is the gravitational constant, we calculate masses of
$M_A=1.4^{+1.2}_{-0.7}$\,\Msol\, and
$M_B=1.5^{+4.4}_{-1.1}$\,\Msol\ for the K-type subgiant and the hot
WD, respectively. We note that the large errors on the masses are mainly
a consequence of large uncertainties of the surface gravities of both
stars. The masses can be constrained more precisely when we compare
their locations in the Hertzsprung Russell diagram (HRD) with
predictions from stellar evolutionary calculations. But this mass
determination also has its limitations, as is outlined in the
following. In the upper panel of Fig.\,\ref{fig:HRD}, we show the
location of \uca in the HRD compared to stellar evolutionary tracks of
\cite{Pietrinferni+2004}. The tracks are calculated for $Z=0.01$
([Fe/H] = 0--0.25~dex) and a helium mass fraction of 0.259 (upper
panel in Fig.\,\ref{fig:HRD}) and suggest a mass of
$M_A=1.2^{+0.7}_{-0.4}$\,\Msol. In the lower panel of
Fig.\,\ref{fig:HRD}, we show the evolutionary tracks of
\cite{Pietrinferni+2004}, which were calculated for a slightly higher
metallicity of $Z=0.03$ ([Fe/H] = 0.26~dex) and a helium mass fraction
of 0.288. These tracks are shifted toward higher effective
temperatures, and thus suggest a higher mass of
$M_A=1.9^{+0.5}_{-0.9}$\,\Msol. This shows that the mass of the
K-type subgiant obtained from the HRD depends significantly on the
chemistry used in the evolutionary calculations. We therefore infer
a possible  mass range of $M_A=0.8-2.4$\,\Msol\ based on these
evolutionary models.

In Fig.\,\ref{fig:HRD2}, we show the location of \ucb in the HRD compared to
evolutionary calculations of H-shell burning post-asymptotic giant
branch (AGB) stars from \cite{MillerBertolami2016}. This suggests a mass of
$M_B=0.54\pm0.02$\,\Msol\ and that the surface gravity of the WD
might be rather at the lower boundary of the error margin (\loggw{7.0\pm0.1})
determined spectroscopically. In gray, a He-shell burning
very late thermal pulse (VLTP) evolutionary track is shown. This track
extends toward higher \Teff\ at the luminosity found for \ucb, which
supports the assumption that the WD is indeed H-rich (higher envelope
mass; much lower mass not possible due to the limited age of the
disk/Milky Way). In Fig.\,\ref{fig:HRD2}, we also indicate the initial
masses assumed for the computation of the evolutionary tracks. This
suggests that if the initial-to-final mass relation (IFMR) assumed in these
calculations are accurate and the system has not interacted in previous
evolutionary phases, the initial mass of the hot WD was about
1\,\Msol,\ or only slightly higher. A low mass of the WD is also supported
by the absence of a PN. The tracks of
\cite{MillerBertolami2016} calculated for $Z=0.01$ predict a post-AGB
age of about 72\,kyrs for the 0.52\,\Msol\ model and only about
10\,kyrs for the 0.56\,\Msol\ ($Z=0.01$) model. This means that if
\ucb should have ejected a PN on the AGB, and we assume that a PN is
typically visible for 30\,kyrs, then, in the case of a low mass post-AGB
star, the PN should have faded away already, while for higher masses,
the PN should still be visible. Radii, luminosities, and masses of \uca
and \ucb are summarized in Table\,\ref{tab:RLM}.

If we assume again that the system has not undergone mass transfer in
its past and the initial mass of the WD was between 1.0 and
1.25\,\Msol, this would imply that the initial mass of \uca is
only slightly lower than this. The total lifetime of stars in that
mass range at solar metallicity stated by \cite{MillerBertolami2016}
lies between 5.3 and 12.5\,Gyrs. Thus, it is entirely possible that
the system indeed belongs to the thick disk, which is generally found
to be exclusively old, meaning older than nine Gyrs
\citep{Feltzing+2009, Kilic+2017, Hayden+2017}. We also note that in
the course of their study of main-sequence turnoff and subgiant stars
from the AMBRE:HARPS survey, \cite{Hayden+2017} find  stars at the
metallicity of \uca\ that have ages between 8 and 10\,Gyrs. However,
since the IFMR is poorly constrained at the low-mass end, and
mass loss on the RGB is not well understood, it is not possible to
give a more precise estimate of the system's age.

\section{Light curve variability}
\label{sect:lightcurve}

\begin{figure}[t]
\centering
\includegraphics[width=\columnwidth]{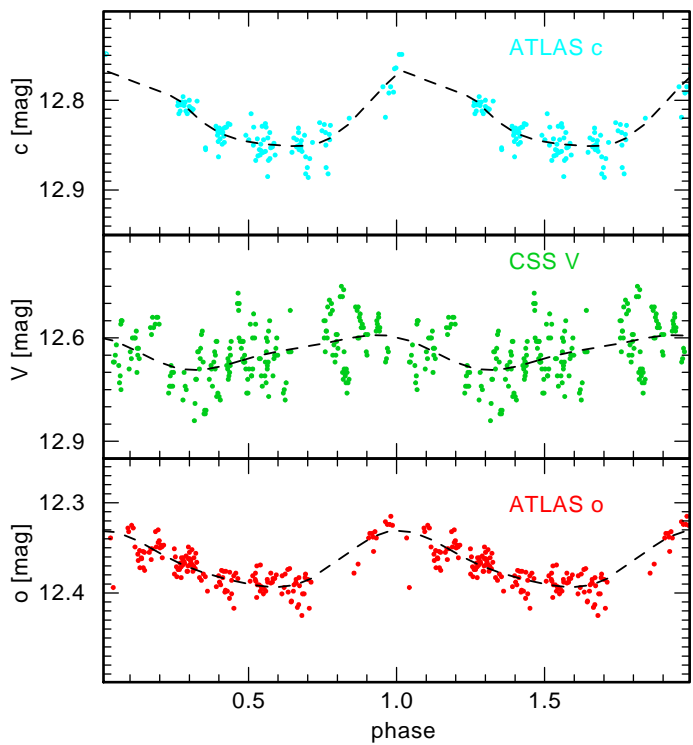}
\caption{ATLAS c-band, CSS V-band, and ATLAS o-band phase-folded light
  curves. The CSS light curve is folded with the 1.97861678\,d period.}
\label{fig:lcs}
\end{figure}

\begin{figure}[t]
\centering
\includegraphics[width=\columnwidth]{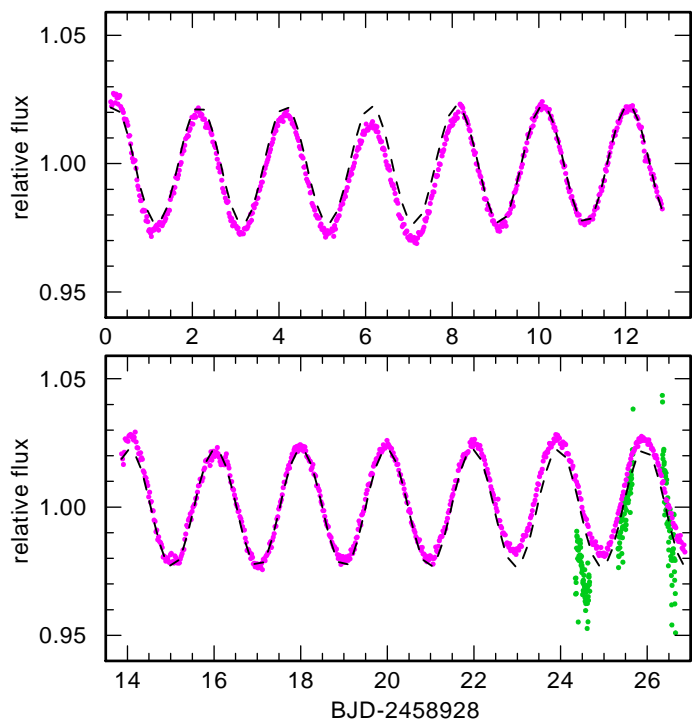}
\caption{TESS (magenta) and T\"ubingen 80\,cm telescope V-band (green) light curves. The V-band
  light curve has been shifted by one period.}
\label{fig:tess}
\end{figure}

We obtained the Catalina Sky Survey (CSS, \citealt{Drake+2009}) DR2
V-band light curve of \uc as well as c- and o-band light curves
(effective wavelengths $0.53\,\mathrm{\mu m}$ and $0.68\,\mathrm{\mu
  m}$, respectively) from the Asteroid Terrestrial-impact Last Alert
System (ATLAS, \citealt{Tonry+2018}) DR1 (Fig.\,\ref{fig:lcs}).

We also observed the light curve of \uc\ in the V-band using the
T\"ubingen 80\,cm f/8 telescope on April 14, 15, and 16
(Fig.\,\ref{fig:tess}). The images were taken with an integration time
of 120\,s using a SBIG STL-1001 CCD camera and a binning of $2\times2$
pixels. Almost every night, data could be obtained over seven
hours. The sky was clear during the first two nights, whereas thin
cirrus was passing throughout the third night. All images were
background and flatfield corrected using the IDL software Time
Resolved Imaging Photometry Package (TRIPP,
\citealt{2003BaltA..12..167S}), which was also used to perform aperture
photometry. The flux error was minimized by testing different aperture
radii with respect to comparison star variances. We note that the
data were obtained under non-photometric conditions, with the weather
conditions being worst in the third night. As there is only one
equally bright star in the field of view, we used only one comparison
star for the first and third nights. During the second night, the observing
conditions were best, so a second comparison star could be used.

In addition, we used \emph{Transiting Exoplanet Survey Satellite} (TESS)
observations of \uc\ (Fig.\,\ref{fig:tess}). The star was in the TESS
field of view during Sector 23 observations. As there is no 2-minute
cadence light curve available, we performed photometry using the
30-minute cadence full-frame images (FFIs) using {\sc lightkurve}
\citep{2018ascl.soft12013L}. We used a relatively small $4\times4$
pixel aperture to minimize contamination from two nearby stars, and
subtracted a background estimated using the same aperture on a nearby
region with no sources.

For the analyses of the light curves, we used the VARTOOLS program
\citep{HartmanBakos2016} to perform a generalized Lomb-Scargle search
\citep{ZechmeisterKuerster2009, Press1992} for
periodic sinusoidal signals. For the ATLAS c- and o-bands, we find the
strongest signal at $P=1.97913923$\,d and $P=1.97899913$\,d, with
associated false alarm probabilities (FAPs) of $10^{-37}$ and
$10^{-28}$, respectively. The derived period from the o-band equals
exactly half the period given in \cite{Heinze+2018}, who classified
the light curve of \uc as  an irregular variable. This class serves in
their catalog as a ``catch-all'' bin for objects that do not seem to fit
into any of their more specific categories.  We also performed a
generalized Lomb-Scargle search on the CSS light curve, and find the
strongest signal at $P=0.21064926$\,d, which, however, has a much
higher FAP of $10^{-13}$. A possible explanation
could be that \uc is indeed an irregular variable and that its
photometric period and amplitude change over time. Thus, in the much
longer CSS light curve (time coverage $\approx 8$\,yrs) no clean
signal can be detected. Another point to notice for the CSS light curve
is that during MJD = 53550 and MJD = 53901, \uc underwent a brightness
variation of 0.7\,mag. Using only CSS data obtained after MJD = 54000,
we repeated the generalized Lomb-Scargle search by looking only for
periods between 1.9 and 2.1\,d. By that, we find a significant peak
at $P=1.97861678$ (FAP $=10^{-6}$), close to the value found for the
ATLAS light curves. In Fig.\,\ref{fig:lcs}, we show the
phase-folded ATLAS c- and o-band and CSS light curves.
The black dashed lines are fits of a harmonic series to the
phase-folded light curves (see Equation (49) in
\citealt{HartmanBakos2016}) and predict peak-to-peak amplitudes of
0.08\,mag, 0.10\,mag, and 0.06\,mag in the V-, c- and o-bands,
respectively. We note, however, that due to the photometric uncertainties
(0.05\,mag for CSS, and 0.01\,mag for the ATLAS light curves), the
differences in the amplitudes might not be real.

For the TESS light curve, we derive a period of $1.98063847$\,d (FAP
$=10^{-755}$). The light curve is shown in Fig.\,\ref{fig:tess}  along
with the V-band light curve obtained with the T\"ubingen 80\,cm
telescope, which is shifted for one period.  We find that the
phases of both light curves match nicely, though the amplitudes
differ.  We note, however, that the amplitudes of both  should be
treated with caution. As mentioned above, the T\"ubingen 80\,cm
telescope light curve was obtained under non-photometric conditions,
and due to the large pixel-size of TESS two neighboring nearby stars
may add additional continuum flux, which in turn may reduce the
amplitude of the TESS light curve. What can also be seen in
Fig.\,\ref{fig:tess} is that the shape of the TESS light curve changes
slightly, which is typical for star spot evolution
\citep{Kovari+2019}.

If we assume the observed period of the light curve variation corresponds
to an orbital period, and a mass of 0.54\,\Msol\ for the WD and 1.0\,\Msol\
of the K-type subgiant (Sect.\,\ref{sect:radii}), then we calculate via
the third Kepler law, a separation of the two stars of $a=7.7$\,\Rsol. In that
case, a strong reflection effect should be noticeable. In such systems, a tidally-locked
cool companion is strongly irradiated by the hot primary, also causing
sinusoidal light curve variations at the orbital period of the system.
Then, however, the amplitudes of the light curve variations should be
increasing toward longer wavelengths, which is not observed. In addition,
we would expect to see all Balmer lines, the CNO line complex around 4560\,\AA,
and possibly also some other lines in emission. These two observational properties
are seen in \object{BE\,UMa}, which also has a \Teff = 105\,000\,K hot primary, an
orbital period of $2.29$\,d, and a distance $a=8.6$\,\Rsol\ from its cool
companion \citep{1994ApJS...94..723F} and in many other reflection effect
systems containing a very hot primary (e.g., \citealt{2004AJ....127.2936S,
  2005MNRAS.359..315E, 2006A&A...448L..25N, 2015ARep...59..199S,
  2016ARep...60..252M}).

Thus, we conclude that the variations are caused by spots on the surface
of the star and that the photometric period corresponds to the rotational
period of \uca.  With a radius of $5.9^{+0.7}_{-0.5}$\,\Rsol\ and
rotation period of 1.98\,d, we calculate a rotational velocity of
$v_{\mathrm{rot}}=151^{+18}_{-13}$\,km\,s$^{-1}$. If we assume that
the mass  is close to 1\,\Msol\ (Sect.\,\ref{sect:radii}), this value
is close to the breakup velocity.\footnote{The breakup velocity
  $v_\mathrm{breakup}=\sqrt{GM/R}$ of a 1.0/2.0\,\Msol\ star with a
  radius of 5.9\,\Rsol\ is 180/254\,km\,s$^{-1}$, respectively.} With
$v \sin(i)=80.76$\,km\,s$^{-1}$ as suggested by the ASPCAP fit of the
APOGEE spectrum, one can estimate that the inclination of rotational
axis is $i=32^{\circ}$.

If we assume a bound rotation (as seen in many RS CVn stars), that the
inclination angle of the rotational axis is perpendicular to the orbital
plane, and the masses given in Tab.\,\ref{tab:RLM}, then RV
semi-amplitudes of 58--105\,km\,s$^{-1}$ and 23--41\,km\,s$^{-1}$
would be expected for the WD and subgiant, respectively.  The latter
is much larger than the standard deviation of the individual RV
measurements of \uca (4.55\,km\,s$^{-1}$, see Sect.\,\ref{sect:distance})
and, thus, again speaks against a close binary system.

\section{Conclusions}
\label{sect:conclusions}

From the SED fit, the fact that we do not find significant RV
variability, that the RVs of \uca and \ucb agree, and that no
reflection effect is noticeable in the light curves and spectra,
we conclude the system is most likely a wide, physically connected
binary system. \uca is a subgiant just about to climb up the RGB,
while \ucb is an extremely hot WD just about to enter the WD cooling
sequence.

Kinematically, and from its height above the Galactic plane, the system
belongs to the Galactic thick disk. This can also explain the low mass
of the WD without the need for binary interactions, as the thick disk
is generally found to contain exclusively old, and by that low-mass
stars.

\subsection{The white dwarf}

We determined the atmospheric parameters of the hot WD in the
binary \uc. Within error limits, the light metal (C, N, O, F, Si, S)
abundances are solar, whereas iron-group elements (Cr, Mn, Fe, Ni) are
slightly enhanced (0.45--0.70 dex oversolar). The H/He ratio cannot be
determined from the available spectra. However, by comparison with
another WD we can conclude that it is likely a DAO white dwarf.
The effective temperature (\Teff = 105\,000\,K) is the same as that of
\egb, a DAO central star of a PN \citep{2018A&A...616A..73W}. 
The metal abundance pattern of \egb\ is 
very similar, and its H/He ratio is solar (Fig.\,\ref{fig:abu}, right
panel). It was concluded that \egb\ is at the wind limit of its
cooling track, shortly before gravitational settling and radiative
acceleration of elements begin to affect the photospheric abundance
pattern. Before the wind limit, a radiation-driven wind prevents
atomic diffusion \citep[e.g.,][]{2000A&A...359.1042U}. Thus, \ucb\ is
probably a DAO white dwarf. The He abundance could be subsolar since
the slightly oversolar heavy-metal abundances indicate that diffusion
began to act. On the other hand, heavy-metal abundances in the K-tpye
subgiant are also oversolar (Fig.\,\ref{fig:abu}, left panel), thus, the element abundance pattern probably
reflects the initial metallicity of both stars.
\ucb\ could be just crossing the wind-limit. Observational evidence
testifies that the wind limit for H-rich WDs is indeed located at
\Teff $\approx 105\,000$\,K
\citep{2018A&A...616A..73W,2019MNRAS.483.5291W} and recent theoretical
work confirms this result \citep{2020A&A...635A.173K}. 

We emphasize, however, that we cannot rule out that the WD is a
helium-rich DO white dwarf. Before the wind limit, DOs can also
exhibit a near-solar metal abundance pattern 
\citep{2017A&A...601A...8W}. To make a conclusive decision, an observation
covering the position of \ion{He}{ii}~1640\,\AA\ is needed. We can
exclude that the WD  belongs to the non-DA PG1159 class whose members
have very abundant C and O \citep{2006PASP..118..183W}.

\egb\ is slightly more massive ($M=0.58^{+0.12}_{-0.04}$\,\Msol,
Fig.\,\ref{fig:HRD2}) than \ucb ($M=0.54\pm0.02$\,\Msol),
and thus \egb\ evolved from the AGB to its current
position more rapidly (post-AGB time is about
$\log(t_{\mathrm{evol}}$\,yr$^{-1})=3.60^{+1.26}_{-0.09}$)
than \ucb ($\log(t_{\mathrm{evol}}$\,yr$^{-1})\approx4.86$). This 
might explain why for \egb\ the PN is still visible, while for \uca
no PN can be detected. 

\subsection{The K subgiant}

We reported the discovery of a time-variable emission of
H\,$\alpha$ as well as emission line cores in the \ion{Ca}{ii} H and K
doublet, which is a strong hint to chromospheric activity.  Further
evidence for chromospheric activity can be seen from the NUV excess.
Furthermore, we found that the object is photometrically variable with
a period of $\approx$2\,d only. The TESS light curve shows small variations in
the shape of light curve. Hence, we conclude that the variations are
caused by (evolving) spots on the surface of \uca. With the radius
from the SED fit, we find that the star is rotating extremely fast
with $v_{\mathrm{rot}}=151^{+18}_{-13}$\,km\,s$^{-1}$. This marks \uca
as one of the most rapidly rotating subgiants known.

The chromospheric activity, NUV excess flux, star spots, and very
rapid rotation is what is also seen in FK Comae Berenices (FK Com)
stars, a small group (only a handful of objects known) of single, rapidly
rotating, and very active GK-type subgiants, which are believed to be
the product of a recent merger \citep{Eggen+Iben1989}. The prototype
\object{FK\,Com} is the undisputed ``king of spin'' amongst these
single red giants with an equatorial spin velocity of
$v_{\mathrm{eq}}=179$\,km\,s$^{-1}$, followed by \object{HD\,199178},
the so called ``spin vice king'' amongst giant stars
\citep{Costa+2015}, with $v_{\mathrm{eq}}=93$\,km\,s$^{-1}$
\citep{Strassmeier2009}.  A related class is that of the more frequently
known RS Canum Venaticorum (RS CVn) variable stars, which are close
binary systems consisting of a chromospherically active subgiant
component, which exhibits brightness variations caused by large cool
spots \citep{Hall1972}, and which is also known to be fast rotating
\citep{Strassmeier2009}. There are also two other hot (pre-) WDs
known, which are in a binary system with chromospherically active and
rapidly rotating cool companions: \object{LW Hya} and \object{LoTr
  5}. The latter is a wide binary in a highly eccentric orbit
($e\approx0.3$, $P=2700$\,d, \citealt{Jones+2017}) consisting of a
hot pre-WD or WD,  and the cool component, \object{IN
  Com}, is a rapidly rotating ($v_{\mathrm{eq}}=95$\,km\,s$^{-1}$,
\citealt{Strassmeier2009}) and magnetically active G5 III giant that
also shows H\,$\alpha$ line-profile variations
\citep{Kovari+2019}. \object{LW Hya} is a resolved binary system
composed of a hot DAO WD (the ionizing star of \object{A\,35},
\citealt{Ziegler+2012, Loebling+2019}) and a magnetically active G8
III-IV-type companion ($v_{\mathrm{eq}}=127$\,km\,s$^{-1}$,
\citealt{Strassmeier2009}).  For \object{LW Hya}, it is, however, not
clear if the system is really a physical binary, due to the
mismatch of the spectroscopic distance of the WD and the optical
parallax. 

The driving mechanism for the chromospheric activity and the reason
for the rapid rotation observed in \uca remains to be explored.  In
main sequence stars of spectral type F, G, K, and early M, a
self-sustaining magnetic dynamo driven by rotation and convection is
believed to be the source of chromospheric activity. Since due to
magnetic braking, the rotational velocity of the star is expected to
decrease with age, the chromospheric activity is thought to decrease
as well. However, if angular momentum is sustained, by tidal
interaction as in the case of short-period binaries (this holds for
the RS CVn stars), or maintained by convection, chromospheric activity
can be preserved \citep{Zhao+2011}.

Besides a recent stellar merger, sudden dredge-up of angular
momentum from the stellar interior \citep{1989ApJ...346..303S}, and
accretion of a substellar companion \citep{1983ApJ...265..972P,
  2012ApJ...757..109C} are considered as possible explanations for the
abnormally high rotation rates of FK\,Com stars. As mentioned before,
the lack of significant RV variations makes the close-binary scenario
for \uc very unlikely. One may therefore speculate that the system
used to be a hierarchical triple system, and that \uca was produced by
the merger of the inner binary, as was suggested for \object{IN Com}
\citep{Jasniewicz+1987, Malasan+1991}.

\subsection{Has \uc an ultra-wide companion?}

Recently, \cite{Tian+2020} reported that based on common Gaia
  parallaxes and proper motions, \uc is in an ultra-wide binary system
  with \object{Gaia DR2 1391040916768689280} (alias SDSS\,J152844.16+430417.1),
  separated by 75.48951\arcsec\ (111\,634\,AU).
  This fainter star has an absolute Gaia G-band magnitude of
  $M_G=6.76$\,mag and a color index of $G_{\rm BP}-G_{\rm RP}=1.26$ \citep{Tian+2020},
  thus it is likely a K-type main-sequence star. That implies a low mass
  ($M\leq0.8$\,\Msol) of the possible ultra-wide companion, which would also be
  in line with the lower limit of the (initial) masses of \uca and \ucb,
  meaning that they indeed could have formed at the same time.

\subsection{Prospects}

A future detailed spectral analysis of \uca, which considers all
spectral observations at once and also considers the flux contribution
of the hot WD in the blue, would be highly desirable. This would help
to better constrain the chemical composition and the mass of the star,
as well as the age of the system. Near-UV spectroscopy could help to
determine the helium abundance of the WD and to investigate the nature
of the NUV excess flux. Long-term, high-resolution monitoring, that
allows RVs to be measured with an accuracy of a few hundred
m\,s$^{-1}$ , will help to constrain the orbital period of the system
and in turn makes it possible to conclude if the system underwent any mass
transfer or wind accretion in its past, and to check if the merger
scenario could be possible. Spectroscopic follow-up of SDSS\,J152844.16+430417.1 would allow us
  to verify if it is an ultra-wide companion by testing if its RV and
  chemical composition agree with \uca.

\begin{acknowledgements} 
We thank the referee, Uli Heber, for his constructive report.
We appreciate useful discussions with M3B and Stephan Geier. We thank
Matti Dorsch and Boris G\"ansicke for pointing out this system to us
and useful discussions. ARM acknowledges support from the MINECO under
the Ram\'on y Cajal programme (RYC-2016-20254) and the AYA2017-86274-P
grant, and the AGAUR grant SGR-661/2017. The TMAD tool
(\url{http://astro.uni-tuebingen.de/~TMAD}) used for this paper was
constructed as part of the activities of the German Astrophysical
Virtual Observatory. This work has made use of data obtained at the
Thai National Observatory on Doi Inthanon, operated by NARIT. Some of
the data presented in this paper were obtained from the Mikulski
Archive for Space Telescopes (MAST). This research has made use of
NASA's Astrophysics Data System and the SIMBAD database, operated at
CDS, Strasbourg, France. This research made use of Lightkurve, a
Python package for Kepler and TESS data analysis (Lightkurve
Collaboration, 2018). This work has made use of data from the European
Space Agency (ESA) mission {\it Gaia}
(\url{https://www.cosmos.esa.int/gaia}), processed by the {\it Gaia}
Data Processing and Analysis Consortium  (DPAC,
\url{https://www.cosmos.esa.int/web/gaia/dpac/consortium}). Funding
for the DPAC has been provided by  national institutions, in
particular the institutions participating in the {\it Gaia}
Multilateral Agreement.  This work has made use of BaSTI web tools.
This work includes data from the Asteroid Terrestrial-impact Last
Alert System (ATLAS) project. ATLAS is primarily funded to search for
near earth asteroids through NASA grants NN12AR55G, 80NSSC18K0284, and
80NSSC18K1575; byproducts of the NEO search include images and
catalogs from the survey area. The ATLAS science products have been
made possible through the contributions of the University of Hawaii
Institute for Astronomy, the Queen's University Belfast, the Space
Telescope Science Institute, and the South African Astronomical
Observatory.  Guoshoujing Telescope (the Large Sky Area Multi-Object
Fiber Spectroscopic Telescope LAMOST) is a National Major Scientific
Project built by the Chinese Academy of Sciences. Funding for the
project has been provided by the National Development and Reform
Commission. LAMOST is operated and managed by the National
Astronomical Observatories, Chinese Academy of Sciences.  Funding for
the Sloan Digital Sky Survey IV has been provided by the Alfred
P. Sloan Foundation, the U.S. Department of Energy Office of Science,
and the Participating Institutions.  SDSS-IV acknowledges support and
resources from the Center for High-Performance Computing at the
University of Utah. The SDSS web site is www.sdss.org.  SDSS-IV is
managed by the Astrophysical Research Consortium for the Participating
Institutions of the SDSS Collaboration including the Brazilian
Participation Group, the Carnegie Institution for Science, Carnegie
Mellon University, the Chilean Participation Group, the French
Participation Group, Harvard-Smithsonian Center for Astrophysics,
Instituto de Astrof\'isica de Canarias, The Johns Hopkins University,
Kavli Institute for the Physics and Mathematics of the Universe (IPMU)
/  University of Tokyo, the Korean Participation Group, Lawrence
Berkeley National Laboratory,  Leibniz Institut f\"ur Astrophysik
Potsdam (AIP), Max-Planck-Institut f\"ur Astronomie (MPIA Heidelberg),
Max-Planck-Institut f\"ur Astrophysik (MPA Garching),
Max-Planck-Institut f\"ur Extraterrestrische Physik (MPE), National
Astronomical Observatories of China, New Mexico State University,  New
York University, University of Notre Dame, Observat\'ario Nacional /
MCTI, The Ohio State University, Pennsylvania State University,
Shanghai Astronomical Observatory, United Kingdom Participation Group,
Universidad Nacional Aut\'onoma de M\'exico, University of Arizona,
University of Colorado Boulder, University of Oxford, University of
Portsmouth, University of Utah, University of Virginia, University of
Washington, University of Wisconsin,  Vanderbilt University, and Yale
University.

\end{acknowledgements}

\bibliographystyle{aa}
\bibliography{aa}

\begin{thebibliography}{111}
\expandafter\ifx\csname natexlab\endcsname\relax\def\natexlab#1{#1}\fi

\bibitem[{{Allard} {et~al.}(2012){Allard}, {Homeier}, \&
  {Freytag}}]{Allard+2012}
{Allard}, F., {Homeier}, D., \& {Freytag}, B. 2012, Philosophical Transactions
  of the Royal Society of London Series A, 370, 2765

\bibitem[{{Andrews} {et~al.}(2015){Andrews}, {Ag{\"u}eros}, {Gianninas},
  {Kilic}, {Dhital}, \& {Anderson}}]{2015ApJ...815...63A}
{Andrews}, J.~J., {Ag{\"u}eros}, M.~A., {Gianninas}, A., {et~al.} 2015, \apj,
  815, 63

\bibitem[{{Asplund} {et~al.}(2009){Asplund}, {Grevesse}, {Sauval}, \&
  {Scott}}]{2009ARA&A..47..481A}
{Asplund}, M., {Grevesse}, N., {Sauval}, A.~J., \& {Scott}, P. 2009, \araa, 47,
  481

\bibitem[{{Bailer-Jones} {et~al.}(2018){Bailer-Jones}, {Rybizki}, {Fouesneau},
  {Mantelet}, \& {Andrae}}]{Bailer-Jones+2018}
{Bailer-Jones}, C.~A.~L., {Rybizki}, J., {Fouesneau}, M., {Mantelet}, G., \&
  {Andrae}, R. 2018, \aj, 156, 58

\bibitem[{{Baxter} {et~al.}(2014){Baxter}, {Dobbie}, {Parker}, {Casewell},
  {Lodieu}, {Burleigh}, {Lawrie}, {K{\"u}lebi}, {Koester}, \& {Holland
  }}]{2014MNRAS.440.3184B}
{Baxter}, R.~B., {Dobbie}, P.~D., {Parker}, Q.~A., {et~al.} 2014, \mnras, 440,
  3184

\bibitem[{{Bianchi} {et~al.}(2014){Bianchi}, {Conti}, \& {Shiao}}]{Bianchi2014}
{Bianchi}, L., {Conti}, A., \& {Shiao}, B. 2014, VizieR Online Data Catalog,
  2335, 0

\bibitem[{{Carlberg} {et~al.}(2012){Carlberg}, {Cunha}, {Smith}, \&
  {Majewski}}]{2012ApJ...757..109C}
{Carlberg}, J.~K., {Cunha}, K., {Smith}, V.~V., \& {Majewski}, S.~R. 2012,
  \apj, 757, 109

\bibitem[{{Casey} {et~al.}(2016){Casey}, {Hogg}, {Ness}, {Rix}, {Ho}, \&
  {Gilmore}}]{Casey+2016}
{Casey}, A.~R., {Hogg}, D.~W., {Ness}, M., {et~al.} 2016, arXiv e-prints,
  arXiv:1603.03040

\bibitem[{{Catal{\'a}n} {et~al.}(2008){Catal{\'a}n}, {Isern},
  {Garc{\'\i}a-Berro}, {Ribas}, {Allende Prieto}, \&
  {Bonanos}}]{2008A&A...477..213C}
{Catal{\'a}n}, S., {Isern}, J., {Garc{\'\i}a-Berro}, E., {et~al.} 2008, \aap,
  477, 213

\bibitem[{{Chen} {et~al.}(2013){Chen}, {Han}, {Deca}, \&
  {Podsiadlowski}}]{2013MNRAS.434..186C}
{Chen}, X., {Han}, Z., {Deca}, J., \& {Podsiadlowski}, P. 2013, \mnras, 434,
  186

\bibitem[{{Costa} {et~al.}(2015){Costa}, {Canto Martins}, {Bravo},
  {Paz-Chinch{\'o}n}, {das Chagas}, {Le{\~a}o}, {Pereira de Oliveira},
  {Rodrigues da Silva}, {Roque}, {de Oliveira}, {Freire da Silva}, \& {De
  Medeiros}}]{Costa+2015}
{Costa}, A.~D., {Canto Martins}, B.~L., {Bravo}, J.~P., {et~al.} 2015, \apjl,
  807, L21

\bibitem[{{Cutri} {et~al.}(2003){Cutri}, {Skrutskie}, {van Dyk}, {Beichman},
  {Carpenter}, {Chester}, {Cambresy}, {Evans}, {Fowler}, {Gizis}, {Howard},
  {Huchra}, {Jarrett}, {Kopan}, {Kirkpatrick}, {Light}, {Marsh}, {McCallon},
  {Schneider}, {Stiening}, {Sykes}, {Weinberg}, {Wheaton}, {Wheelock}, \&
  {Zacarias}}]{Cutri+2003}
{Cutri}, R.~M., {Skrutskie}, M.~F., {van Dyk}, S., {et~al.} 2003, VizieR Online
  Data Catalog, II/246

\bibitem[{{Davis} {et~al.}(2010){Davis}, {Kolb}, \&
  {Willems}}]{2010MNRAS.403..179D}
{Davis}, P.~J., {Kolb}, U., \& {Willems}, B. 2010, \mnras, 403, 179

\bibitem[{{Dixon} {et~al.}(2020){Dixon}, {Tayar}, \& {Stassun}}]{Dixon+2020}
{Dixon}, D., {Tayar}, J., \& {Stassun}, K.~G. 2020, arXiv e-prints,
  arXiv:2005.00577

\bibitem[{{Drake} {et~al.}(2009){Drake}, {Djorgovski}, {Mahabal}, {Beshore},
  {Larson}, {Graham}, {Williams}, {Christensen}, {Catelan}, {Boattini},
  {Gibbs}, {Hill}, \& {Kowalski}}]{Drake+2009}
{Drake}, A.~J., {Djorgovski}, S.~G., {Mahabal}, A., {et~al.} 2009, \apj, 696,
  870

\bibitem[{{Duch{\^e}ne} \& {Kraus}(2013)}]{Duchene+Kraus2013}
{Duch{\^e}ne}, G. \& {Kraus}, A. 2013, \araa, 51, 269

\bibitem[{{Eggen} \& {Iben}(1989)}]{Eggen+Iben1989}
{Eggen}, O.~J. \& {Iben}, Icko, J. 1989, \aj, 97, 431

\bibitem[{{Exter} {et~al.}(2005){Exter}, {Pollacco}, {Maxted}, {Napiwotzki}, \&
  {Bell}}]{2005MNRAS.359..315E}
{Exter}, K.~M., {Pollacco}, D.~L., {Maxted}, P.~F.~L., {Napiwotzki}, R., \&
  {Bell}, S.~A. 2005, \mnras, 359, 315

\bibitem[{{Feltzing} \& {Bensby}(2009)}]{Feltzing+2009}
{Feltzing}, S. \& {Bensby}, T. 2009, in IAU Symposium, Vol. 258, The Ages of
  Stars, ed. E.~E. {Mamajek}, D.~R. {Soderblom}, \& R.~F.~G. {Wyse}, 23--30

\bibitem[{{Ferguson} \& {James}(1994)}]{1994ApJS...94..723F}
{Ferguson}, D.~H. \& {James}, T.~A. 1994, \apjs, 94, 723

\bibitem[{{Fitzpatrick}(1999)}]{fitzpatrick1999}
{Fitzpatrick}, E.~L. 1999, \pasp, 111, 63

\bibitem[{{Fuhrmann}(1998)}]{Fuhrmann1998}
{Fuhrmann}, K. 1998, \aap, 338, 161

\bibitem[{{Garc{\'\i}a P{\'e}rez} {et~al.}(2016){Garc{\'\i}a P{\'e}rez},
  {Allende Prieto}, {Holtzman}, {Shetrone}, {M{\'e}sz{\'a}ros}, {Bizyaev},
  {Carrera}, {Cunha}, {Garc{\'\i}a-Hern{\'a}ndez}, {Johnson}, {Majewski},
  {Nidever}, {Schiavon}, {Shane}, {Smith}, {Sobeck}, {Troup}, {Zamora},
  {Weinberg}, {Bovy}, {Eisenstein}, {Feuillet}, {Frinchaboy}, {Hayden},
  {Hearty}, {Nguyen}, {O'Connell}, {Pinsonneault}, {Wilson}, \&
  {Zasowski}}]{ASPCAP+2016}
{Garc{\'\i}a P{\'e}rez}, A.~E., {Allende Prieto}, C., {Holtzman}, J.~A.,
  {et~al.} 2016, \aj, 151, 144

\bibitem[{{Geier} {et~al.}(2015){Geier}, {Kupfer}, {Heber}, {Schaffenroth},
  {Barlow}, {{\O}stensen}, {O'Toole}, {Ziegerer}, {Heuser}, {Maxted},
  {G{\"a}nsicke}, {Marsh}, {Napiwotzki}, {Br{\"u}nner}, {Schindewolf}, \&
  {Niederhofer}}]{Geier+2015}
{Geier}, S., {Kupfer}, T., {Heber}, U., {et~al.} 2015, \aap, 577, A26

\bibitem[{{Geier} {et~al.}(2013){Geier}, {Marsh}, {Wang}, {Dunlap}, {Barlow},
  {Schaffenroth}, {Chen}, {Irrgang}, {Maxted}, {Ziegerer}, {Kupfer},
  {Miszalski}, {Heber}, {Han}, {Shporer}, {Telting}, {G{\"a}nsicke},
  {{\O}stensen}, {O'Toole}, \& {Napiwotzki}}]{2013A&A...554A..54G}
{Geier}, S., {Marsh}, T.~R., {Wang}, B., {et~al.} 2013, \aap, 554, A54

\bibitem[{{Gianninas} {et~al.}(2015){Gianninas}, {Kilic}, {Brown}, {Canton}, \&
  {Kenyon}}]{Gianninas+2015}
{Gianninas}, A., {Kilic}, M., {Brown}, W.~R., {Canton}, P., \& {Kenyon}, S.~J.
  2015, \apj, 812, 167

\bibitem[{{Gray} \& {Corbally}(2009)}]{GrayCorbally2009}
{Gray}, R.~O. \& {Corbally}, Christopher, J. 2009, {Stellar Spectral
  Classification}

\bibitem[{{Hall}(1972)}]{Hall1972}
{Hall}, D.~S. 1972, \pasp, 84, 323

\bibitem[{{Han} {et~al.}(2002){Han}, {Podsiadlowski}, {Maxted}, {Marsh}, \&
  {Ivanova}}]{2002MNRAS.336..449H}
{Han}, Z., {Podsiadlowski}, P., {Maxted}, P.~F.~L., {Marsh}, T.~R., \&
  {Ivanova}, N. 2002, \mnras, 336, 449

\bibitem[{{Hartman} \& {Bakos}(2016)}]{HartmanBakos2016}
{Hartman}, J.~D. \& {Bakos}, G.~{\'A}. 2016, Astronomy and Computing, 17, 1

\bibitem[{{Hayden} {et~al.}(2017){Hayden}, {Recio-Blanco}, {de Laverny},
  {Mikolaitis}, \& {Worley}}]{Hayden+2017}
{Hayden}, M.~R., {Recio-Blanco}, A., {de Laverny}, P., {Mikolaitis}, S., \&
  {Worley}, C.~C. 2017, \aap, 608, L1

\bibitem[{{Heinze} {et~al.}(2018){Heinze}, {Tonry}, {Denneau}, {Flewelling},
  {Stalder}, {Rest}, {Smith}, {Smartt}, \& {Weiland}}]{Heinze+2018}
{Heinze}, A.~N., {Tonry}, J.~L., {Denneau}, L., {et~al.} 2018, \aj, 156, 241

\bibitem[{{Henden} {et~al.}(2015){Henden}, {Levine}, {Terrell}, \&
  {Welch}}]{Henden+2015}
{Henden}, A.~A., {Levine}, S., {Terrell}, D., \& {Welch}, D.~L. 2015, in
  American Astronomical Society Meeting Abstracts, Vol. 225, American
  Astronomical Society Meeting Abstracts \#225, 336.16

\bibitem[{{Ho} {et~al.}(2017){Ho}, {Ness}, {Hogg}, {Rix}, {Liu}, {Yang},
  {Zhang}, {Hou}, \& {Wang}}]{Ho+2017}
{Ho}, A. Y.~Q., {Ness}, M.~K., {Hogg}, D.~W., {et~al.} 2017, \apj, 836, 5

\bibitem[{{Holberg} \& {Bergeron}(2006)}]{HolbergBergeron2006}
{Holberg}, J.~B. \& {Bergeron}, P. 2006, \aj, 132, 1221

\bibitem[{{Holberg} {et~al.}(2013){Holberg}, {Oswalt}, {Sion}, {Barstow}, \&
  {Burleigh}}]{2013MNRAS.435.2077H}
{Holberg}, J.~B., {Oswalt}, T.~D., {Sion}, E.~M., {Barstow}, M.~A., \&
  {Burleigh}, M.~R. 2013, \mnras, 435, 2077

\bibitem[{{Jasniewicz} {et~al.}(1987){Jasniewicz}, {Duquennoy}, \&
  {Acker}}]{Jasniewicz+1987}
{Jasniewicz}, G., {Duquennoy}, A., \& {Acker}, A. 1987, \aap, 180, 145

\bibitem[{{Jones} {et~al.}(2017){Jones}, {Van Winckel}, {Aller}, {Exter}, \&
  {De Marco}}]{Jones+2017}
{Jones}, D., {Van Winckel}, H., {Aller}, A., {Exter}, K., \& {De Marco}, O.
  2017, \aap, 600, L9

\bibitem[{{K{\H{o}}v{\'a}ri} {et~al.}(2019){K{\H{o}}v{\'a}ri}, {Strassmeier},
  {Ol{\'a}h}, {Kriskovics}, {Vida}, {Carroll}, {Granzer}, {Ilyin}, {Jurcsik},
  {K{\H{o}}v{\'a}ri}, \& {Weber}}]{Kovari+2019}
{K{\H{o}}v{\'a}ri}, Z., {Strassmeier}, K.~G., {Ol{\'a}h}, K., {et~al.} 2019,
  \aap, 624, A83

\bibitem[{{Kilic} {et~al.}(2017){Kilic}, {Munn}, {Harris}, {von Hippel},
  {Liebert}, {Williams}, {Jeffery}, \& {DeGennaro}}]{Kilic+2017}
{Kilic}, M., {Munn}, J.~A., {Harris}, H.~C., {et~al.} 2017, \apj, 837, 162

\bibitem[{{Kordopatis} {et~al.}(2013){Kordopatis}, {Gilmore}, {Steinmetz},
  {Boeche}, {Seabroke}, {Siebert}, {Zwitter}, {Binney}, {de Laverny},
  {Recio-Blanco}, {Williams}, {Piffl}, {Enke}, {Roeser}, {Bijaoui}, {Wyse},
  {Freeman}, {Munari}, {Carrillo}, {Anguiano}, {Burton}, {Campbell}, {Cass},
  {Fiegert}, {Hartley}, {Parker}, {Reid}, {Ritter}, {Russell}, {Stupar},
  {Watson}, {Bienaym{\'e}}, {Bland -Hawthorn}, {Gerhard}, {Gibson}, {Grebel},
  {Helmi}, {Navarro}, {Conrad}, {Famaey}, {Faure}, {Just}, {Kos},
  {Matijevi{\v{c}}}, {McMillan}, {Minchev}, {Scholz}, {Sharma}, {Siviero}, {de
  Boer}, \& {{\v{Z}}erjal}}]{2013AJ....146..134K}
{Kordopatis}, G., {Gilmore}, G., {Steinmetz}, M., {et~al.} 2013, \aj, 146, 134

\bibitem[{{Kordopatis} {et~al.}(2011){Kordopatis}, {Recio-Blanco}, {de
  Laverny}, {Gilmore}, {Hill}, {Wyse}, {Helmi}, {Bijaoui}, {Zoccali}, \&
  {Bienaym{\'e}}}]{Kordopatis+2011}
{Kordopatis}, G., {Recio-Blanco}, A., {de Laverny}, P., {et~al.} 2011, \aap,
  535, A107

\bibitem[{{Kordopatis} {et~al.}(2015){Kordopatis}, {Wyse}, {Gilmore},
  {Recio-Blanco}, {de Laverny}, {Hill}, {Adibekyan}, {Heiter}, {Minchev},
  {Famaey}, {Bensby}, {Feltzing}, {Guiglion}, {Korn}, {Mikolaitis},
  {Schultheis}, {Vallenari}, {Bayo}, {Carraro}, {Flaccomio}, {Franciosini},
  {Hourihane}, {Jofr{\'e}}, {Koposov}, {Lardo}, {Lewis}, {Lind}, {Magrini},
  {Morbidelli}, {Pancino}, {Randich}, {Sacco}, {Worley}, \&
  {Zaggia}}]{Kordopatis+2015}
{Kordopatis}, G., {Wyse}, R.~F.~G., {Gilmore}, G., {et~al.} 2015, \aap, 582,
  A122

\bibitem[{{Krti{\v{c}}ka} {et~al.}(2020){Krti{\v{c}}ka}, {Kub{\'a}t}, \&
  {Krti{\v{c}}kov{\'a}}}]{2020A&A...635A.173K}
{Krti{\v{c}}ka}, J., {Kub{\'a}t}, J., \& {Krti{\v{c}}kov{\'a}}, I. 2020, \aap,
  635, A173

\bibitem[{{Lallement} {et~al.}(2018){Lallement}, {Capitanio}, {Ruiz-Dern},
  {Danielski}, {Babusiaux}, {Vergely}, {Elyajouri}, {Arenou}, \&
  {Leclerc}}]{Lallement+2018}
{Lallement}, R., {Capitanio}, L., {Ruiz-Dern}, L., {et~al.} 2018, \aap, 616,
  A132

\bibitem[{{Lightkurve Collaboration} {et~al.}(2018){Lightkurve Collaboration},
  {Cardoso}, {Hedges}, {Gully-Santiago}, {Saunders}, {Cody}, {Barclay}, {Hall},
  {Sagear}, {Turtelboom}, {Zhang}, {Tzanidakis}, {Mighell}, {Coughlin}, {Bell},
  {Berta-Thompson}, {Williams}, {Dotson}, \& {Barentsen}}]{2018ascl.soft12013L}
{Lightkurve Collaboration}, {Cardoso}, J. V. d. M.~a., {Hedges}, C., {et~al.}
  2018, {Lightkurve: Kepler and TESS time series analysis in Python}

\bibitem[{{L{\"o}bling} {et~al.}(2019){L{\"o}bling}, {Maney}, {Rauch},
  {Quinet}, {Gamrath}, {Kruk}, \& {Werner}}]{Loebling+2019}
{L{\"o}bling}, L., {Maney}, M.~A., {Rauch}, T., {et~al.} 2019, \mnras, 2831

\bibitem[{{Luo} {et~al.}(2016){Luo}, {Zhao}, {Zhao}, {Deng}, {Liu}, {Jing},
  {Wang}, {Zhang}, {Shi}, {Cui}, {Chu}, {Li}, {Bai}, {Wu}, {Cai}, {Cao}, {Cao},
  {Carlin}, {Chen}, {Chen}, {Chen}, {Chen}, {Chen}, {Chen}, {Chen},
  {Christlieb}, {Chu}, {Cui}, {Dong}, {Du}, {Fan}, {Feng}, {Fu}, {Gao}, {Gong},
  {Gu}, {Guo}, {Han}, {He}, {Hou}, {Hou}, {Hou}, {Hu}, {Hu}, {Hu}, {Huo},
  {Jia}, {Jiang}, {Jiang}, {Jiang}, {Jin}, {Kong}, {Kong}, {Lei}, {Li}, {Li},
  {Li}, {Li}, {Li}, {Li}, {Li}, {Li}, {Li}, {Li}, {Li}, {Li}, {Liang}, {Lin},
  {Liu}, {Liu}, {Liu}, {Liu}, {Lu}, {Luo}, {Mao}, {Newberg}, {Ni}, {Qi}, {Qi},
  {Shen}, {Shi}, {Song}, {Song}, {Su}, {Su}, {Tang}, {Tao}, {Tian}, {Wang},
  {Wang}, {Wang}, {Wang}, {Wang}, {Wang}, {Wang}, {Wang}, {Wang}, {Wang},
  {Wang}, {Wang}, {Wang}, {Wang}, {Wang}, {Wang}, {Wang}, {Wang}, {Wang},
  {Wang}, {Wei}, {Wei}, {Wu}, {Wu}, {Wu}, {Wu}, {Xing}, {Xu}, {Xu}, {Xu},
  {Yan}, {Yang}, {Yang}, {Yang}, {Yang}, {Yao}, {Yu}, {Yuan}, {Yuan}, {Yuan},
  {Yuan}, {Zhai}, {Zhang}, {Zhang}, {Zhang}, {Zhang}, {Zhang}, {Zhang},
  {Zhang}, {Zhang}, {Zhao}, {Zhou}, {Zhou}, {Zhu}, {Zhu}, {Zou}, \&
  {Zuo}}]{Luo+2016}
{Luo}, A.~L., {Zhao}, Y.~H., {Zhao}, G., {et~al.} 2016, VizieR Online Data
  Catalog, V/149

\bibitem[{{Luo} {et~al.}(2019){Luo}, {Zhao}, {Zhao}, \& {et al.}}]{Luo+2019}
{Luo}, A.~L., {Zhao}, Y.~H., {Zhao}, G., \& {et al.} 2019, VizieR Online Data
  Catalog, V/164

\bibitem[{{Majewski} {et~al.}(2017){Majewski}, {Schiavon}, {Frinchaboy},
  {Allende Prieto}, {Barkhouser}, {Bizyaev}, {Blank}, {Brunner}, {Burton},
  {Carrera}, {Chojnowski}, {Cunha}, {Epstein}, {Fitzgerald}, {Garc{\'\i}a
  P{\'e}rez}, {Hearty}, {Henderson}, {Holtzman}, {Johnson}, {Lam}, {Lawler},
  {Maseman}, {M{\'e}sz{\'a}ros}, {Nelson}, {Nguyen}, {Nidever}, {Pinsonneault},
  {Shetrone}, {Smee}, {Smith}, {Stolberg}, {Skrutskie}, {Walker}, {Wilson},
  {Zasowski}, {Anders}, {Basu}, {Beland}, {Blanton}, {Bovy}, {Brownstein},
  {Carlberg}, {Chaplin}, {Chiappini}, {Eisenstein}, {Elsworth}, {Feuillet},
  {Fleming}, {Galbraith-Frew}, {Garc{\'\i}a}, {Garc{\'\i}a-Hern{\'a}ndez},
  {Gillespie}, {Girardi}, {Gunn}, {Hasselquist}, {Hayden}, {Hekker}, {Ivans},
  {Kinemuchi}, {Klaene}, {Mahadevan}, {Mathur}, {Mosser}, {Muna}, {Munn},
  {Nichol}, {O'Connell}, {Parejko}, {Robin}, {Rocha-Pinto}, {Schultheis},
  {Serenelli}, {Shane}, {Silva Aguirre}, {Sobeck}, {Thompson}, {Troup},
  {Weinberg}, \& {Zamora}}]{Apogee2017}
{Majewski}, S.~R., {Schiavon}, R.~P., {Frinchaboy}, P.~M., {et~al.} 2017, \aj,
  154, 94

\bibitem[{{Malasan} {et~al.}(1991){Malasan}, {Yamasaki}, \&
  {Kondo}}]{Malasan+1991}
{Malasan}, H.~L., {Yamasaki}, A., \& {Kondo}, M. 1991, \aj, 101, 2131

\bibitem[{{Marsh}(1989)}]{1989PASP..101.1032M}
{Marsh}, T.~R. 1989, \pasp, 101, 1032

\bibitem[{{Maxted} {et~al.}(2001){Maxted}, {Heber}, {Marsh}, \&
  {North}}]{Maxted2001}
{Maxted}, P.~F.~L., {Heber}, U., {Marsh}, T.~R., \& {North}, R.~C. 2001,
  \mnras, 326, 1391

\bibitem[{{Miller Bertolami}(2016)}]{MillerBertolami2016}
{Miller Bertolami}, M.~M. 2016, \aap, 588, A25

\bibitem[{{Mitrofanova} {et~al.}(2016){Mitrofanova}, {Shimansky}, {Borisov},
  {Spiridonova}, \& {Gabdeev}}]{2016ARep...60..252M}
{Mitrofanova}, A.~A., {Shimansky}, V.~V., {Borisov}, N.~V., {Spiridonova},
  O.~I., \& {Gabdeev}, M.~M. 2016, Astronomy Reports, 60, 252

\bibitem[{{Moe} {et~al.}(2019){Moe}, {Kratter}, \&
  {Badenes}}]{2019ApJ...875...61M}
{Moe}, M., {Kratter}, K.~M., \& {Badenes}, C. 2019, \apj, 875, 61

\bibitem[{{Nagel} {et~al.}(2006){Nagel}, {Schuh}, {Kusterer}, {Stahn},
  {H{\"u}gelmeyer}, {Dreizler}, {G{\"a}nsicke}, \&
  {Schreiber}}]{2006A&A...448L..25N}
{Nagel}, T., {Schuh}, S., {Kusterer}, D.~J., {et~al.} 2006, \aap, 448, L25

\bibitem[{{Napiwotzki} {et~al.}(2001){Napiwotzki}, {Christlieb}, {Drechsel},
  {Hagen}, {Heber}, {Homeier}, {Karl}, {Koester}, {Leibundgut}, {Marsh},
  {Moehler}, {Nelemans}, {Pauli}, {Reimers}, {Renzini}, \&
  {Yungelson}}]{Napiwotzkietal2001}
{Napiwotzki}, R., {Christlieb}, N., {Drechsel}, H., {et~al.} 2001,
  Astronomische Nachrichten, 322, 411

\bibitem[{{Nebot G{\'o}mez-Mor{\'a}n} {et~al.}(2011){Nebot
  G{\'o}mez-Mor{\'a}n}, {G{\"a}nsicke}, {Schreiber}, {Rebassa-Mansergas},
  {Schwope}, {Southworth}, {Aungwerojwit}, {Bothe}, {Davis}, {Kolb},
  {M{\"u}ller}, {Papadaki}, {Pyrzas}, {Rabitz}, {Rodr{\'\i}guez-Gil},
  {Schmidtobreick}, {Schwarz}, {Tappert}, {Toloza}, {Vogel}, \&
  {Zorotovic}}]{2011A&A...536A..43N}
{Nebot G{\'o}mez-Mor{\'a}n}, A., {G{\"a}nsicke}, B.~T., {Schreiber}, M.~R.,
  {et~al.} 2011, \aap, 536, A43

\bibitem[{{Ness} {et~al.}(2015){Ness}, {Hogg}, {Rix}, {Ho}, \&
  {Zasowski}}]{Ness+2015}
{Ness}, M., {Hogg}, D.~W., {Rix}, H.~W., {Ho}, A. Y.~Q., \& {Zasowski}, G.
  2015, \apj, 808, 16

\bibitem[{{Nie} {et~al.}(2012){Nie}, {Wood}, \&
  {Nicholls}}]{2012MNRAS.423.2764N}
{Nie}, J.~D., {Wood}, P.~R., \& {Nicholls}, C.~P. 2012, \mnras, 423, 2764

\bibitem[{{Parsons} {et~al.}(2016){Parsons}, {Rebassa-Mansergas}, {Schreiber},
  {G{\"a}nsicke}, {Zorotovic}, \& {Ren}}]{2016MNRAS.463.2125P}
{Parsons}, S.~G., {Rebassa-Mansergas}, A., {Schreiber}, M.~R., {et~al.} 2016,
  \mnras, 463, 2125

\bibitem[{{Pastetter} \& {Ritter}(1989)}]{1989A&A...214..186P}
{Pastetter}, L. \& {Ritter}, H. 1989, \aap, 214, 186

\bibitem[{{Peterson} {et~al.}(1983){Peterson}, {Tarbell}, \&
  {Carney}}]{1983ApJ...265..972P}
{Peterson}, R.~C., {Tarbell}, T.~D., \& {Carney}, B.~W. 1983, \apj, 265, 972

\bibitem[{{Pietrinferni} {et~al.}(2004){Pietrinferni}, {Cassisi}, {Salaris}, \&
  {Castelli}}]{Pietrinferni+2004}
{Pietrinferni}, A., {Cassisi}, S., {Salaris}, M., \& {Castelli}, F. 2004, \apj,
  612, 168

\bibitem[{{Press} {et~al.}(1992){Press}, {Teukolsky}, {Vetterling}, \&
  {Flannery}}]{Press1992}
{Press}, W.~H., {Teukolsky}, S.~A., {Vetterling}, W.~T., \& {Flannery}, B.~P.
  1992, {Numerical recipes in C. The art of scientific computing}

\bibitem[{{Rauch} {et~al.}(2007){Rauch}, {Ziegler}, {Werner}, {Kruk},
  {Oliveira}, {Vande Putte}, {Mignani}, \& {Kerber}}]{2007A&A...470..317R}
{Rauch}, T., {Ziegler}, M., {Werner}, K., {et~al.} 2007, \aap, 470, 317

\bibitem[{{Rebassa-Mansergas} {et~al.}(2017){Rebassa-Mansergas}, {Ren},
  {Irawati}, {Garc{\'\i}a-Berro}, {Parsons}, {Schreiber}, {G{\"a}nsicke},
  {Rodr{\'\i}guez-Gil}, {Liu}, {Manser}, {Nevado}, {Jim{\'e}nez-Ibarra},
  {Costero}, {Echevarr{\'\i}a}, {Michel}, {Zorotovic}, {Hollands}, {Han},
  {Luo}, {Villaver}, \& {Kong}}]{2017MNRAS.472.4193R}
{Rebassa-Mansergas}, A., {Ren}, J.~J., {Irawati}, P., {et~al.} 2017, \mnras,
  472, 4193

\bibitem[{{Rebassa-Mansergas} {et~al.}(2016){Rebassa-Mansergas}, {Ren},
  {Parsons}, {G{\"a}nsicke}, {Schreiber}, {Garc{\'\i}a-Berro}, {Liu}, \&
  {Koester}}]{2016MNRAS.458.3808R}
{Rebassa-Mansergas}, A., {Ren}, J.~J., {Parsons}, S.~G., {et~al.} 2016, \mnras,
  458, 3808

\bibitem[{{Rebassa-Mansergas} {et~al.}(2019){Rebassa-Mansergas}, {Toonen},
  {Korol}, \& {Torres}}]{Rebassa-Mansergas+2019}
{Rebassa-Mansergas}, A., {Toonen}, S., {Korol}, V., \& {Torres}, S. 2019,
  \mnras, 482, 3656

\bibitem[{{Recio-Blanco} {et~al.}(2014){Recio-Blanco}, {de Laverny},
  {Kordopatis}, {Helmi}, {Hill}, {Gilmore}, {Wyse}, {Adibekyan}, {Randich},
  {Asplund}, {Feltzing}, {Jeffries}, {Micela}, {Vallenari}, {Alfaro}, {Allende
  Prieto}, {Bensby}, {Bragaglia}, {Flaccomio}, {Koposov}, {Korn}, {Lanzafame},
  {Pancino}, {Smiljanic}, {Jackson}, {Lewis}, {Magrini}, {Morbidelli},
  {Prisinzano}, {Sacco}, {Worley}, {Hourihane}, {Bergemann}, {Costado},
  {Heiter}, {Joffre}, {Lardo}, {Lind}, \& {Maiorca}}]{Recio-Blanco+2014}
{Recio-Blanco}, A., {de Laverny}, P., {Kordopatis}, G., {et~al.} 2014, \aap,
  567, A5

\bibitem[{{Reindl} {et~al.}(2016){Reindl}, {Geier}, {Kupfer}, {Bloemen},
  {Schaffenroth}, {Heber}, {Barlow}, \& {{\O}stensen}}]{Reindl+2016}
{Reindl}, N., {Geier}, S., {Kupfer}, T., {et~al.} 2016, \aap, 587, A101

\bibitem[{{Reindl} {et~al.}(2020){Reindl}, {Schaffenroth}, {Miller Bertolami},
  {Geier}, {Finch}, {Barstow}, {Casewell}, \& {Taubenberger}}]{Reindl+2020}
{Reindl}, N., {Schaffenroth}, V., {Miller Bertolami}, M.~M., {et~al.} 2020,
  \aap, 638, A93

\bibitem[{{Ritter}(1986)}]{1986A&A...168..105R}
{Ritter}, H. 1986, \aap, 168, 105

\bibitem[{{Santander-Garc{\'{\i}}a} {et~al.}(2015){Santander-Garc{\'{\i}}a},
  {Rodr{\'{\i}}guez-Gil}, {Corradi}, {Jones}, {Miszalski}, {Boffin},
  {Rubio-D{\'{\i}}ez}, \& {Kotze}}]{SG+2015}
{Santander-Garc{\'{\i}}a}, M., {Rodr{\'{\i}}guez-Gil}, P., {Corradi}, R.~L.~M.,
  {et~al.} 2015, \nat, 519, 63

\bibitem[{{Schlafly} \& {Finkbeiner}(2011)}]{Schlafly2011}
{Schlafly}, E.~F. \& {Finkbeiner}, D.~P. 2011, \apj, 737, 103

\bibitem[{{Schlegel} {et~al.}(1998){Schlegel}, {Finkbeiner}, \&
  {Davis}}]{Schlegel1998}
{Schlegel}, D.~J., {Finkbeiner}, D.~P., \& {Davis}, M. 1998, \apj, 500, 525

\bibitem[{{Schreiber} {et~al.}(2010){Schreiber}, {G{\"a}nsicke},
  {Rebassa-Mansergas}, {Nebot Gomez-Moran}, {Southworth}, {Schwope},
  {M{\"u}ller}, {Papadaki}, {Pyrzas}, {Rabitz}, {Rodr{\'\i}guez-Gil},
  {Schmidtobreick}, {Schwarz}, {Tappert}, {Toloza}, {Vogel}, \&
  {Zorotovic}}]{2010A&A...513L...7S}
{Schreiber}, M.~R., {G{\"a}nsicke}, B.~T., {Rebassa-Mansergas}, A., {et~al.}
  2010, \aap, 513, L7

\bibitem[{{Schuh} {et~al.}(2003){Schuh}, {Dreizler}, {Deetjen}, \&
  {G{\"o}hler}}]{2003BaltA..12..167S}
{Schuh}, S.~L., {Dreizler}, S., {Deetjen}, J.~L., \& {G{\"o}hler}, E. 2003,
  Baltic Astronomy, 12, 167

\bibitem[{{Shimansky} {et~al.}(2015){Shimansky}, {Borisov}, {Nurtdinova},
  {Solovyeva}, {Sakhibullin}, \& {Spiridonova}}]{2015ARep...59..199S}
{Shimansky}, V.~V., {Borisov}, N.~V., {Nurtdinova}, D.~N., {et~al.} 2015,
  Astronomy Reports, 59, 199

\bibitem[{{Silvestri} {et~al.}(2005){Silvestri}, {Hawley}, \&
  {Oswalt}}]{2005AJ....129.2428S}
{Silvestri}, N.~M., {Hawley}, S.~L., \& {Oswalt}, T.~D. 2005, \aj, 129, 2428

\bibitem[{{Simon} \& {Drake}(1989)}]{1989ApJ...346..303S}
{Simon}, T. \& {Drake}, S.~A. 1989, \apj, 346, 303

\bibitem[{{Sing} {et~al.}(2004){Sing}, {Holberg}, {Burleigh}, {Good},
  {Barstow}, {Oswalt}, {Howell}, {Brinkworth}, {Rudkin}, {Johnston}, \&
  {Rafferty}}]{2004AJ....127.2936S}
{Sing}, D.~K., {Holberg}, J.~B., {Burleigh}, M.~R., {et~al.} 2004, \aj, 127,
  2936

\bibitem[{{Stelzer} {et~al.}(2016){Stelzer}, {Damasso}, {Scholz}, \&
  {Matt}}]{2016MNRAS.463.1844S}
{Stelzer}, B., {Damasso}, M., {Scholz}, A., \& {Matt}, S.~P. 2016, \mnras, 463,
  1844

\bibitem[{{Stelzer} {et~al.}(2013){Stelzer}, {Marino}, {Micela},
  {L{\'o}pez-Santiago}, \& {Liefke}}]{2013MNRAS.431.2063S}
{Stelzer}, B., {Marino}, A., {Micela}, G., {L{\'o}pez-Santiago}, J., \&
  {Liefke}, C. 2013, \mnras, 431, 2063

\bibitem[{{Strassmeier}(2009)}]{Strassmeier2009}
{Strassmeier}, K.~G. 2009, \aapr, 17, 251

\bibitem[{{Tian} {et~al.}(2020){Tian}, {El-Badry}, {Rix}, \&
  {Gould}}]{Tian+2020}
{Tian}, H.-J., {El-Badry}, K., {Rix}, H.-W., \& {Gould}, A. 2020, \apjs, 246, 4

\bibitem[{{Tian} {et~al.}(2015){Tian}, {Liu}, {Carlin}, {Zhao}, {Chen}, {Wu},
  {Li}, {Hou}, \& {Zhang}}]{Tian+2015}
{Tian}, H.-J., {Liu}, C., {Carlin}, J.~L., {et~al.} 2015, \apj, 809, 145

\bibitem[{{Tonry} {et~al.}(2018){Tonry}, {Denneau}, {Heinze}, {Stalder},
  {Smith}, {Smartt}, {Stubbs}, {Weiland}, \& {Rest}}]{Tonry+2018}
{Tonry}, J.~L., {Denneau}, L., {Heinze}, A.~N., {et~al.} 2018, \pasp, 130,
  064505

\bibitem[{{Unglaub} \& {Bues}(2000)}]{2000A&A...359.1042U}
{Unglaub}, K. \& {Bues}, I. 2000, \aap, 359, 1042

\bibitem[{{Van der Swaelmen} {et~al.}(2017){Van der Swaelmen}, {Boffin},
  {Jorissen}, \& {Van Eck}}]{2017A&A...597A..68V}
{Van der Swaelmen}, M., {Boffin}, H.~M.~J., {Jorissen}, A., \& {Van Eck}, S.
  2017, \aap, 597, A68

\bibitem[{{Van Winckel}(2018)}]{2018arXiv180900871V}
{Van Winckel}, H. 2018, arXiv e-prints, arXiv:1809.00871

\bibitem[{{Werner} {et~al.}(2003){Werner}, {Deetjen}, {Dreizler}, {Nagel},
  {Rauch}, \& {Schuh}}]{2003ASPC..288...31W}
{Werner}, K., {Deetjen}, J.~L., {Dreizler}, S., {et~al.} 2003, in Astronomical
  Society of the Pacific Conference Series, Vol. 288, Stellar Atmosphere
  Modeling, ed. I.~{Hubeny}, D.~{Mihalas}, \& K.~{Werner}, 31

\bibitem[{{Werner} \& {Dreizler}(1999)}]{1999JCoAM.109...65W}
{Werner}, K. \& {Dreizler}, S. 1999, Journal of Computational and Applied
  Mathematics, 109, 65

\bibitem[{{Werner} {et~al.}(2012){Werner}, {Dreizler}, \& {Rauch}}]{tmap2012}
{Werner}, K., {Dreizler}, S., \& {Rauch}, T. 2012, {TMAP: T{\"u}bingen NLTE
  Model-Atmosphere Package}, Astrophysics Source Code Library [record
  ascl:1212.015]

\bibitem[{{Werner} \& {Herwig}(2006)}]{2006PASP..118..183W}
{Werner}, K. \& {Herwig}, F. 2006, \pasp, 118, 183

\bibitem[{{Werner} {et~al.}(2015){Werner}, {Rauch}, \&
  {Kruk}}]{2015A&A...582A..94W}
{Werner}, K., {Rauch}, T., \& {Kruk}, J.~W. 2015, \aap, 582, A94

\bibitem[{{Werner} {et~al.}(2017){Werner}, {Rauch}, \&
  {Kruk}}]{2017A&A...601A...8W}
{Werner}, K., {Rauch}, T., \& {Kruk}, J.~W. 2017, \aap, 601, A8

\bibitem[{{Werner} {et~al.}(2018{\natexlab{a}}){Werner}, {Rauch}, \&
  {Kruk}}]{2018A&A...609A.107W}
{Werner}, K., {Rauch}, T., \& {Kruk}, J.~W. 2018{\natexlab{a}}, \aap, 609, A107

\bibitem[{{Werner} {et~al.}(2018{\natexlab{b}}){Werner}, {Rauch}, \&
  {Kruk}}]{2018A&A...616A..73W}
{Werner}, K., {Rauch}, T., \& {Kruk}, J.~W. 2018{\natexlab{b}}, \aap, 616, A73

\bibitem[{{Werner} {et~al.}(2019){Werner}, {Rauch}, \&
  {Reindl}}]{2019MNRAS.483.5291W}
{Werner}, K., {Rauch}, T., \& {Reindl}, N. 2019, \mnras, 483, 5291

\bibitem[{{Willems} \& {Kolb}(2004)}]{2004A&A...419.1057W}
{Willems}, B. \& {Kolb}, U. 2004, \aap, 419, 1057

\bibitem[{{Wilson}(1963)}]{Wilson1963}
{Wilson}, O.~C. 1963, \apj, 138, 832

\bibitem[{{Wilson}(1968)}]{Wilson1968}
{Wilson}, O.~C. 1968, \apj, 153, 221

\bibitem[{{Wright} {et~al.}(2010){Wright}, {Eisenhardt}, {Mainzer}, {Ressler},
  {Cutri}, {Jarrett}, {Kirkpatrick}, {Padgett}, {McMillan}, {Skrutskie},
  {Stanford}, {Cohen}, {Walker}, {Mather}, {Leisawitz}, {Gautier}, {McLean},
  {Benford}, {Lonsdale}, {Blain}, {Mendez}, {Irace}, {Duval}, {Liu}, {Royer},
  {Heinrichsen}, {Howard}, {Shannon}, {Kendall}, {Walsh}, {Larsen}, {Cardon},
  {Schick}, {Schwalm}, {Abid}, {Fabinsky}, {Naes}, \& {Tsai}}]{Wright+2010}
{Wright}, E.~L., {Eisenhardt}, P. R.~M., {Mainzer}, A.~K., {et~al.} 2010, \aj,
  140, 1868

\bibitem[{{Zechmeister} \& {K{\"u}rster}(2009)}]{ZechmeisterKuerster2009}
{Zechmeister}, M. \& {K{\"u}rster}, M. 2009, \aap, 496, 577

\bibitem[{{Zhao} {et~al.}(2012{\natexlab{a}}){Zhao}, {Zhao}, {Chu}, {Jing}, \&
  {Deng}}]{Zhao+2012}
{Zhao}, G., {Zhao}, Y.-H., {Chu}, Y.-Q., {Jing}, Y.-P., \& {Deng}, L.-C.
  2012{\natexlab{a}}, Research in Astronomy and Astrophysics, 12, 723

\bibitem[{{Zhao} {et~al.}(2011){Zhao}, {Oswalt}, {Rudkin}, {Zhao}, \&
  {Chen}}]{Zhao+2011}
{Zhao}, J.~K., {Oswalt}, T.~D., {Rudkin}, M., {Zhao}, G., \& {Chen}, Y.~Q.
  2011, \aj, 141, 107

\bibitem[{{Zhao} {et~al.}(2012{\natexlab{b}}){Zhao}, {Oswalt}, {Willson},
  {Wang}, \& {Zhao}}]{2012ApJ...746..144Z}
{Zhao}, J.~K., {Oswalt}, T.~D., {Willson}, L.~A., {Wang}, Q., \& {Zhao}, G.
  2012{\natexlab{b}}, \apj, 746, 144

\bibitem[{{Ziegler} {et~al.}(2012){Ziegler}, {Rauch}, {Werner}, {K{\"o}ppen},
  \& {Kruk}}]{Ziegler+2012}
{Ziegler}, M., {Rauch}, T., {Werner}, K., {K{\"o}ppen}, J., \& {Kruk}, J.~W.
  2012, \aap, 548, A109

\bibitem[{{Zorotovic} {et~al.}(2011){Zorotovic}, {Schreiber}, {G{\"a}nsicke},
  {Rebassa-Mansergas}, {Nebot G{\'o}mez-Mor{\'a}n}, {Southworth}, {Schwope},
  {Pyrzas}, {Rodr{\'\i}guez-Gil}, {Schmidtobreick}, {Schwarz}, {Tappert},
  {Toloza}, \& {Vogt}}]{2011A&A...536L...3Z}
{Zorotovic}, M., {Schreiber}, M.~R., {G{\"a}nsicke}, B.~T., {et~al.} 2011,
  \aap, 536, L3

\end{thebibliography}

\end{document}